\pdfoutput=0
\documentclass[twocolumn]{emulateapj}
\usepackage{graphicx}
\usepackage{natbib}
\usepackage{multirow}
\usepackage{textcomp}

\newcommand{\kms}{km~s$^{-1}$}
\newcommand{\hii}{HII }
\newcommand{\am}{NH$_{3}$}

\newcommand{\cm}{cm$^{-3}$}

\begin{document}

\slugcomment{}

\title{Detection of Widespread Hot Ammonia in the Galactic Center}

\author{E.A.C. Mills, M. R.  Morris}
\affil{Department of Physics and Astronomy, University of California, Los Angeles}
\email{millsb@astro.ucla.edu}

\begin{abstract}

We present the detection of metastable inversion lines of ammonia (\am) from energy levels high above the ground state. We detect these lines in both emission and absorption toward fifteen of seventeen positions in the central 300 parsecs of the Galaxy. In total, we observe seven metastable transitions of  \am: (8,8), 
(9,9), (10,10), (11,11), (12,12),  (13,13) and (15,15),  with energies (in Kelvins) ranging from 680 to 2200 K. We also mapped emission from  \am\, (8,8) and (9,9) in two clouds in the Sgr A complex (M-0.02-0.07 and M-0.13-0.08), and we find that the line emission is concentrated toward the the dense centers of these molecular clouds. The rotational temperatures derived from the metastable lines toward M-0.02-0.07 and M-0.13-0.08 and an additional cloud (M0.25+0.01) range from 350 to 450 K. Similarly highly-excited lines of \am\, have previously been observed toward Sgr B2, where gas with kinetic temperatures of $\sim$ 600 K had been inferred. Our observations show that the existence of a hot molecular gas component is not unique to Sgr B2, but rather appears common to many Galactic center molecular clouds.
In M-0.02-0.07, we find that the hot \am\, contributes $\sim$ 10\% of the cloud's total \am\, column density, and further, that the hot \am\, in this cloud arises in gas which is extended or uniformly distributed on $\sim$ 10 arcsecond scales. We discuss the implications of these constraints upon the nature of this hot gas component. In addition to the detection of hot metastable \am\, line emission, we also detect for the first time emission from nonmetastable inversion transitions of \am\, in both M-0.02-0.07 and M-0.13-0.08. \end{abstract}

\keywords{Galactic Center}

\section{Introduction}

The physical conditions in the molecular gas of the central 300 parsecs of the Galaxy (the Central Molecular Zone or CMZ) differ significantly from conditions in the molecular gas in the rest of the Galactic disk. Molecular gas in the CMZ exists primarily in giant molecular clouds of sizes 15 to 50 pc, and masses 10$^4-10^6$ M$_{\odot}$. The average densities in these clouds (10$^4-10^5$ \cm) are a few orders of magnitude above those typically found in Galactic disk clouds.  Gas in Galactic center giant molecular clouds (GCMCs) is also more turbulent, as indicated by typical linewidths of $20-30$ km s$^{-1}$ \citep{Bally88,Morris96}. The gas in these clouds, most of which are not actively forming stars, is further measured to be warmer \citep[T $\sim$ 25 - 200 K, ][]{Gusten81, Huttem93b,Ao13} than gas in quiescent molecular clouds in the disk, which is generally measured to have T $\sim$ 5 - 10 K.

The observed temperature structure of CMZ clouds is complex. \cite{Huttem93b} used ammonia to measure two temperature components, 25 K and 200 K, with the 25 K component attributed to gas with densities of $n \sim$ 10$^5$ \cm\, and the 200 K component attributed to less dense gas of n $\sim$ 10$^4$ \cm. \cite{Ao13} confirm via observations of formaldehyde that the dense gas has temperatures up to at least 100 K. \cite{RF01} also observe a hot (T$\sim400-600$ K) and dense ($n\lesssim10^6$\cm) molecular gas envelope in CMZ clouds in rotational transitions of H$_2$. In Sgr B2, the densest and most massive molecular cloud in the CMZ, which is actively forming massive stars, an extremely hot (T$\sim$ 700 K) component of gas is observed in absorption lines of \am\, \citep{Mauers86, Huttem93, Huttem95, Cecca02, Wilson06}. Not only is hot gas associated with molecular clouds in the CMZ, there is also evidence for a diffuse ($n \sim50-200$ \cm), warm (T = 250-400 K) molecular gas component throughout the CMZ, inferred from absorption line measurements of H$_3^+$ toward numerous lines of sight in the CMZ \citep{Oka05,Goto08,Goto11} and measurements of H$_3$O$^+$ and para-H$_2$O toward Sgr B2 and Sgr A \citep{Lis10b,Lis12,Sonnen13}.

 The ammonia molecule (\am) is a very useful thermometer for dense molecular gas. Its relatively low dipole moment ($\mu$ = 1.47 D) means that its metastable levels can relatively quickly come to thermal equilibrium with the surrounding gas via collisions with H$_2$ for densities of n $\ge$ $10^4$ \cm.  \am\, also has a symmetric top structure which gives rise to inversion doublets as the nitrogen atom tunnels back and forth through the potential barrier presented by the plane of the hydrogen atoms. The non-metastable ($J \ne K$) levels decay downward quickly (A $\sim10^{-2}$ sec$^{-1}$ ), populating the metastable $J=K$ levels. These latter levels cannot radiatively decay by allowed transitions \citep[except for a slow octopole transition;]{Oka71}. The $\Delta K = \pm$3 collisional transitions are similarly very slow \citep[A $\sim 10^{-9}$ sec$^{-1}$,][]{Cheung69}, so that the rotational temperature derived from the ratio of column densities of metastable levels is close to but less than the kinetic temperature, T$_{\mathrm K}$\citep{Morris73, Danby88}. Because the decay time of the non-metastable levels is so much shorter, the rotational temperatures derived from the column densities of non-metastable levels with $\Delta K$ = 0 are not sensitive to the kinetic temperature, but are instead sensitive to the local excitation conditions, reflecting the gas density and/or the intensity of the far-infrared radiation field. 

Using measurements of highly excited metastable inversion lines of \am, we find evidence for a T$\gtrsim$ 400 K gas component, similar to that observed toward Sgr B2 \citep{Wilson06}, but which appears common to dense molecular clouds in the CMZ.  We present spectra of  \am\, (8,8) and (9,9) lines toward a sample of 15 molecular clouds in the CMZ, as well as spectra of higher-excitation \am\, lines toward three of these clouds. Additionally, we map the distribution of \am\, (8,8) and (9,9) emission in the molecular clouds near Sgr A: M-0.02-0.07 and M-0.13-0.08, and we detect non-metastable ammonia emission in these clouds for the first time. Based on these observations, we discuss our constraints upon the density and column density of this gas, and its relation to previously-observed hot molecular gas components in the CMZ. 

\section{Observations and Data Reduction}

\begin{figure*}
\includegraphics[scale=0.35]{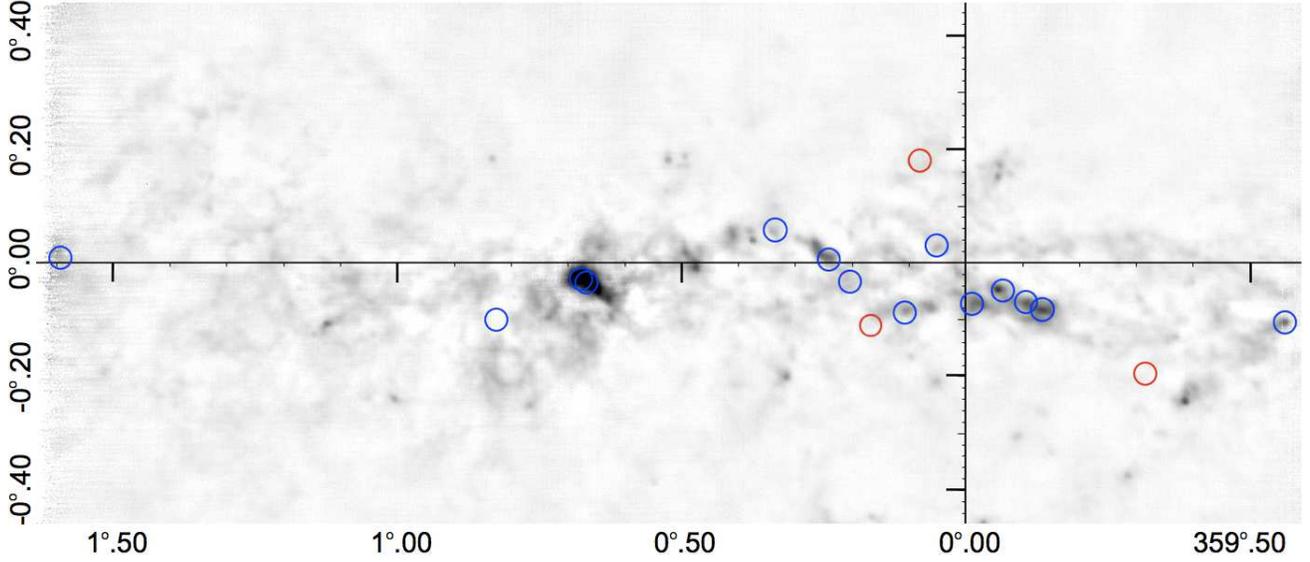}
\caption{ Positions of observed \am\, spectra overlaid on a Bolocam 1.1 mm image of the CMZ \citep{Bally10}.  Blue circles represent positions where \am\, (9,9) is detected (in either emission or absorption); red circles represent non-detections of \am\, (9,9).}
\label{map}
\end{figure*}

The observations presented in this paper were primarily made with the 100 m Green Bank telescope (GBT) of the National Radio Astronomy Observatory\footnotemark[1]\footnotetext[1]{The National Radio Astronomy Observatory is a facility of the National Science Foundation operated under cooperative agreement by Associated Universities, Inc.} during the periods of May 12- 29, 2009,  November 12, 2010; January 3 and 23, April 15, and Dec 24, 2011. Single pointing measurements of the (8,8) and (9,9) metastable \am\, inversion lines were made toward a sample of 17 positions in 15 clouds (Table \ref{Sources}; Figure \ref{map}). For three of these positions (M-0.02-0.07, M-0.13-0.08, M0.25+0.01) we additionally observed  the (10,10), (11,11), (12,12), (13,13) and (15,15) lines. We also observed the (2,1), (3,2) and (4,3) non-metastable inversion lines in M-0.02-0.07 and M-0.13-0.08. Frequencies of all the observed transitions are given in Table \ref{ammonia}.  The spectral resolution of the observations was 390.625 KHz, or 3.1 to 4.4 \kms over the observed range of frequencies, sufficient to resolve lines with intrinsic widths of 15 to 30 \kms.

The metastable \am\, inversion transitions were made using one beam from the dual-beam Ka-band receiver, and the non-metastable observations were made using the K-band receiver (decommissioned in 2011). We employed a position-switching technique, with an offset position of RA = 17h46m00s  -$28\degr13'57''$, for clouds east of ($l=0.1\degr$), and an offset position of RA = 17h45m59.9s, Dec. =  -29$\degr 16' 47''$ for more westerly clouds. 

\subsection{GBT Mapping}

In addition to spectra from the pointed observations described above, we also mapped emission from the \am\, (8,8) and (9,9) lines simultaneously over two $5'\times 5'$ fields toward M-0.02-0.07 and M-0.13-0.08. Toward M-0.02-0.07, we additionally mapped emission from the (1,1), (2,2), (3,3), (4,4) and (6,6) line over the same field.  The maps were made using an on-the-fly mapping technique, alternately scanning along both Right Ascension and Declination. After every two scans, an emission-free off position was observed. The data were gridded in AIPS using the routine SDGRD.  

\subsection{VLA Mapping}

Observations of  the \am\, (9,9) line in the M-0.02-0.07 cloud were also made with the VLA in the DnC configuration in January 2012.  The phase calibrator for these observations was J1744-3116, and the flux calibrator was 3C286. The total integration time on this target was 31 minutes. We mapped emission from this line in two overlapping fields centered on R.A., decl. (J2000): ($17^{\mathrm{h}}45^{\mathrm{m}}53^{\mathrm{s}}.266$, $-28\degr59'12''.61$) and ($17^{\mathrm{h}}45^{\mathrm{m}}51^{\mathrm{s}}.245$, $-28\degr59'52''.65$), with a primary beam size of $1.8'$. The spectral resolution of these observations was similar to that of the GBT, 250 kHz or 2.7 \kms. The data were cleaned using the multi-scale CLEAN algorithm in the CASA reduction package\footnotemark[1]\footnotetext[1]{http://casa.nrao.edu/}, resulting in a synthesized clean beam FWHM of $2''.3\times 2''.4$. The noise in the final image is 0.25 Jy beam$^{-1}$.

\subsection{GBT Calibration}

Using the GBTIDL\footnotemark[2]\footnotetext[2]{http://gbtidl.nrao.edu/} reduction and analysis software, we corrected the antenna temperature of the observed targets for the frequency-dependent opacity at the observed elevation. Given the latitude of the GBT, the elevation at which the Galactic center targets were observed was uniformly low, ranging from $10\degr$ to $23\degr$. The atmospheric opacity estimate was taken from a GBT archive of frequency-dependent opacities calculated from the weather conditions at the time of the observations. We then fit for and removed the local baseline fluctuations around each line. 

We next used observations of the flux calibrator 3C286 to more accurately determine the relative amplitude calibration of the data, which is otherwise limited to $10-15\%$ accuracy by temporal fluctuations in the noise diode. The amplitude of the noise diode will also vary with frequency, so this relative calibration is important for an accurate comparison of line intensities across a range of frequencies. The expected flux density (S$_\nu$) of 3C286 at the line frequency is determined by interpolating the data of \cite{Ott94}. The theoretical aperture efficiency as a function of frequency for the GBT (Ron Maddalena, personal communication) is given by:

	\begin{equation} \eta_{\mathrm{eff,theor}} =  2.0\, \mathrm{exp}\left(-\left[0.00922\hspace{0.1cm} \nu\left(\mathrm{GHz}\right)\right]^2\right)
	\label{eqb}
	\end{equation}
	
The measured aperture efficiency for 3C286 can be calculated as:

	\begin{equation} \eta_{\mathrm{eff,meas}} = \mathrm{T}_{A}\, \mathrm{exp}\left(\tau_{\mathrm atm} / \mathrm{sin}\, \theta \right) / \textrm{S}_\nu,
	\label{eqc}
	\end{equation}

\noindent where $\theta$ is the elevation at which the flux calibrator was observed,  and $\tau_{\mathrm atm}$ is the atmospheric opacity estimate calculated from the archived weather conditions at the time of the observations. 3C286 is observed at a higher elevation than the Galactic center, however, as with the Galactic center sources, the resulting differences in the atmospheric path are accounted for by correcting the antenna temperature in the above equation for the frequency-dependent opacity at the observed elevation. 

We then apply an amplitude correction to the data, that is the ratio of the theoretical to measured aperture efficiency. Our measured amplitude corrections vary both with frequency and between observing runs, and range from 0.7 to 1.3, a slightly larger correction than the stated 10-15\% variation attributable to the noise diodes. Finally, assuming that the observed emission is extended over an area larger than the telescope beam (see Section \ref{distribution}), we also apply a correction for the main-beam efficiency, to convert the measured antenna temperatures to main-beam brightness temperatures. For the GBT, this correction is 1.32 at all frequencies. 

We estimate the uncertainty of our relative amplitude calibration to be $<5\%$, based on the RMS fluctuations in the observed spectrum of the phase calibrator.

\section{Results and Analysis}

\subsection{Detection of the metastable  \am\, (9,9) through (15,15) lines}
\begin{figure*}
\includegraphics[scale=0.4]{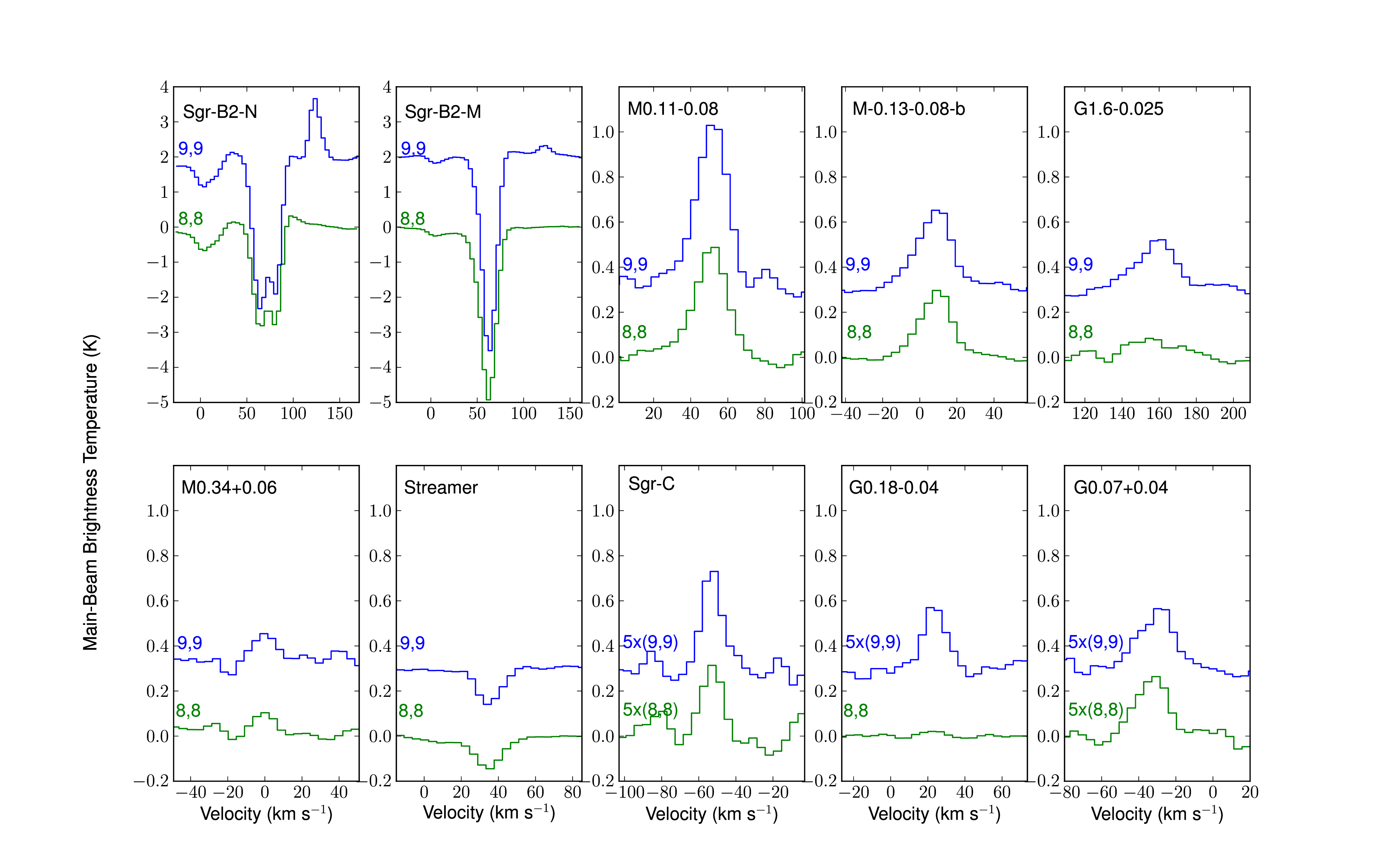}
\caption{ Spectra of \am\, (8,8) and (9,9) toward the 9 positions where at least \am\, (9,9) is detected. In the spectra of \am\, (9,9) toward Sgr B2, the emission line to the right of the \am\, absorption feature is the $J_K$= 13$_2$-13$_1$ line of CH$_3$OH, with a rest frequency of 27.47253 GHz.}
\label{detect}
\end{figure*}

From our survey of 17 positions toward 15 giant molecular clouds we detect the \am\, (9,9) metastable inversion line toward thirteen positions, all but two of which we also simultaneously detect in the \am\, (8,8) line. The locations of the 13 positions for which we detect the (9,9) line are shown in blue in Figure \ref{map}. We detect \am\, (9,9) in emission toward ten of these positions, and in absorption toward three of them (Figure \ref{detect}). We also marginally ($\sigma \lesssim 3$) detect emission from the (8,8) and (9,9) lines toward two additional positions.

\begin{figure*}
\includegraphics[scale=0.4]{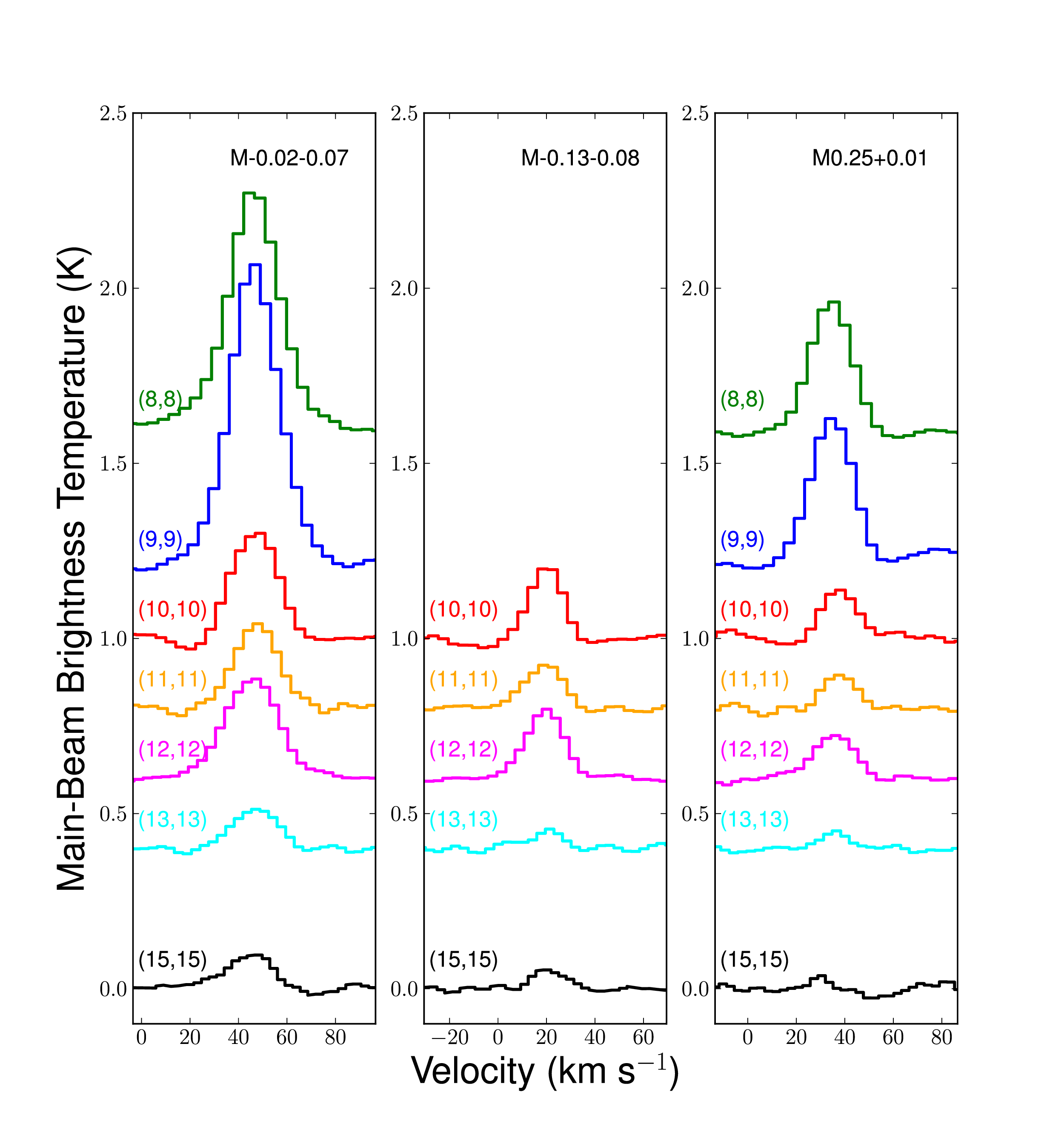}
\caption{ Spectra of \am\, (8,8) through (15,15) toward the three clouds for which we measure the rotational temperature.}
\label{meta}
\end{figure*}

The strongest emission from these highly-excited lines of \am\, is toward M-0.02-0.07. For this cloud and two other of the strongest detections, M-0.13-0.08 and M0.25+0.01, we performed follow-up observations of the \am\, (10,10), (11,11), (12,12), (13,13) and (15,15) lines. We detect all of these lines except for the (15,15) line in M0.25+0.01, for which we have only an upper limit (Figure \ref{meta}). The parameters of all of the observed lines of \am\, are reported in Table \ref{results}. We derive temperatures for these three clouds in Section \ref{RT}.

\begin{figure*}
\includegraphics[scale=0.3, angle=270]{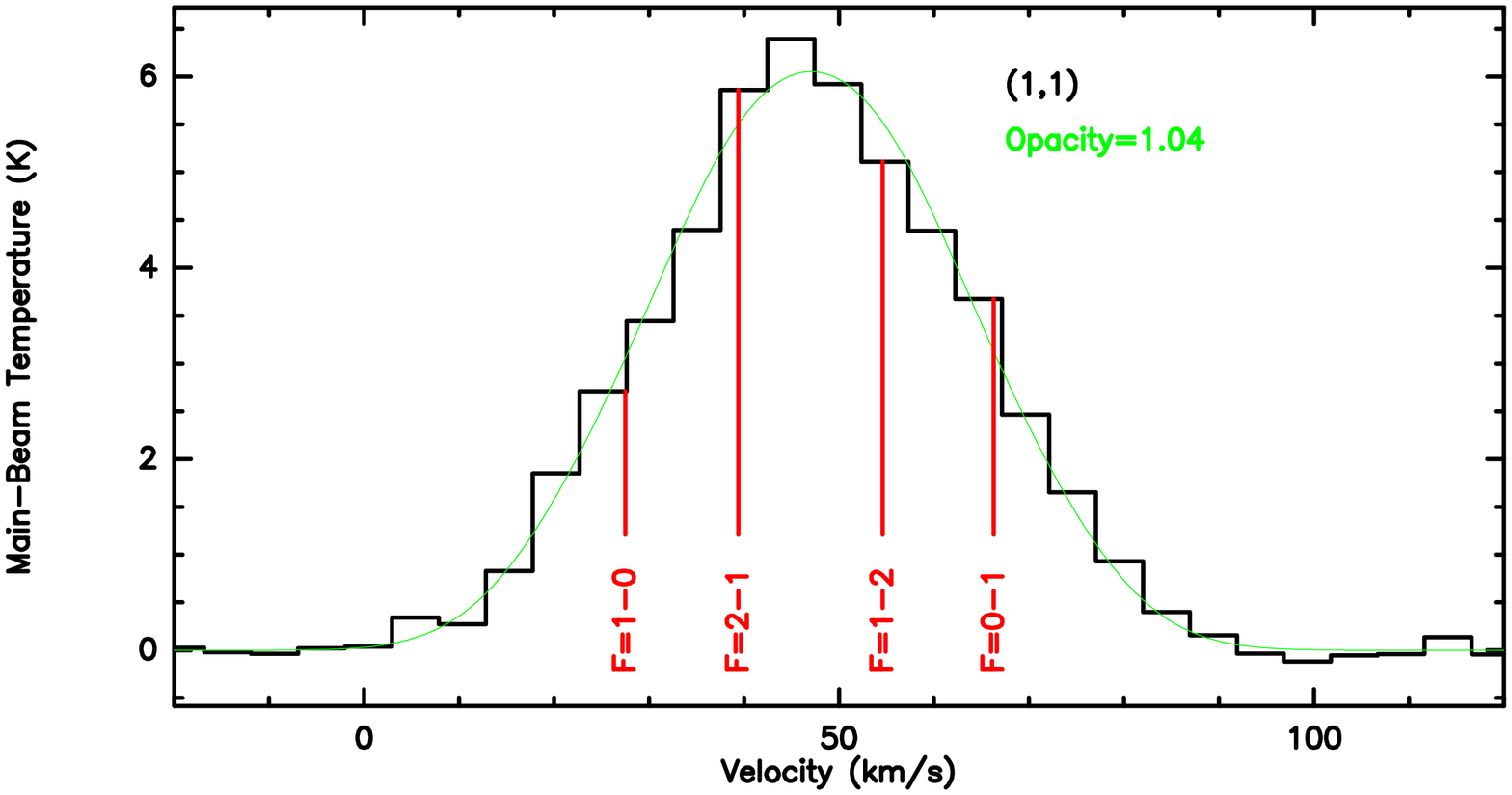}
\includegraphics[scale=0.3, angle=270]{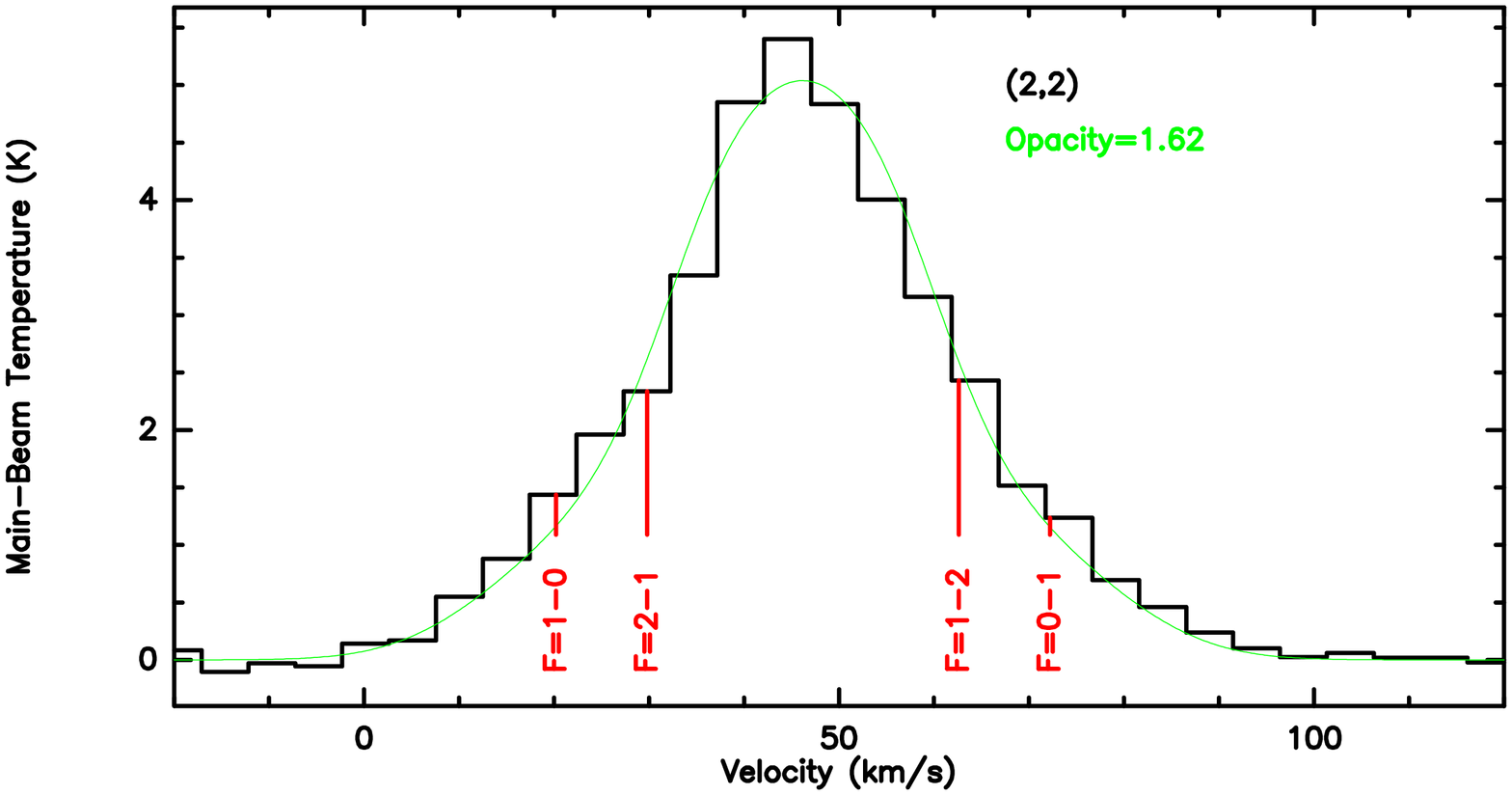}
\caption{ Fits to the hyperfine structure of the two lowest metastable \am\, lines toward M-0.02-0.07. Although the individual hyperfine line components are not resolved, by assuming an intrinsic width we are able fit to the strength of the satellite lines and thus constrain the line opacities.  Here, we have assumed the line width is  22.7 \kms, equal to that of the (3,3), (4,4) and (6,6) lines, for which the hyperfine components are too faint to affect the line profile. The locations of the individual satellite lines are indicated in red.}
\label{hyperfig}
\end{figure*}

\begin{figure*}
\includegraphics[scale=0.4, angle=270]{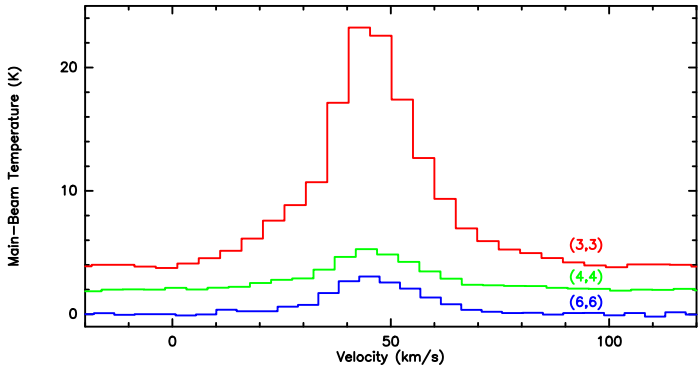}
\caption{Line profiles of the metastable \am\, (3,3) through (6,6) transitions extracted from a map of M-0.02-0.07 at the position of the pointed, higher-excitation line observations shown in Figure \ref{meta}.}
\label{lowfig}
\end{figure*}

\am\, (1,1), (2,2), (3,3), (4,4) and (6,6) spectra from maps of M-0.02-0.07 at the same positions as the higher-excitation lines are also shown in Figures \ref{hyperfig} and \ref{lowfig}.
Parameters for all lines except the (1,1) and (2,2) lines (which have hyperfine structure, and are reported separately in Table \ref{hyperfine}) are also reported in Table \ref{results}. 

\subsection{Nondetections}

\begin{figure*}
\includegraphics[scale=0.4]{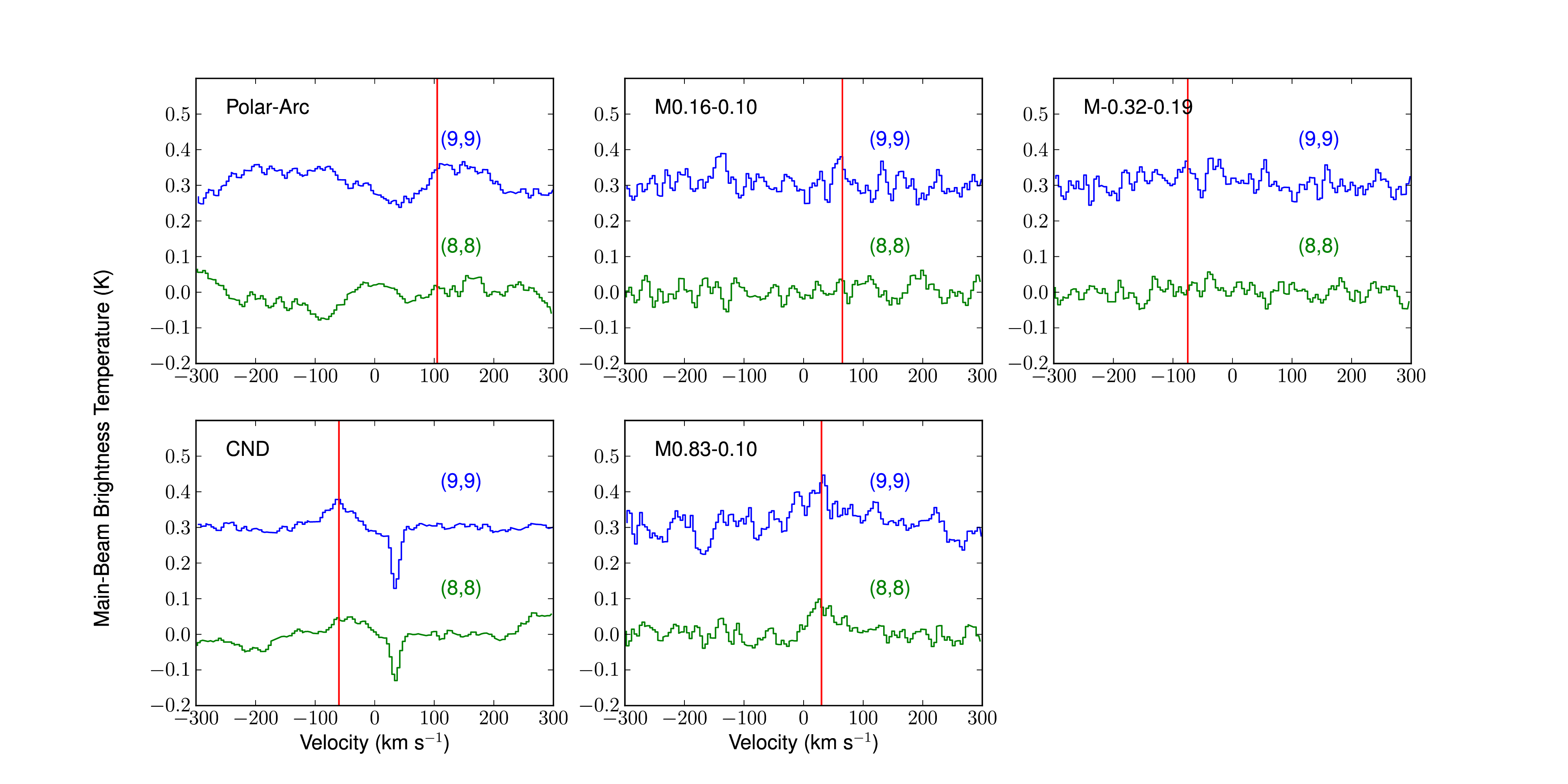}
\caption{ {\bf Top row:} Observations of \am\, (8,8) and (9,9) toward the 3 positions where neither line is detected. The central source velocity is shown as a red line in each plot. {\bf Bottom row:} Observations of  \am\, (8,8) and (9,9)  toward two positions where these lines are marginally detected.  The CND and Polar Arc were observed during more optimal weather conditions on a different date than the other sources, resulting in lower system temperatures and a lower noise level. The absorption line at 20 \kms in the spectrum of the CND is the same as that shown in Figure \ref{detect}, and is from the Southern Streamer feature} %for all of these spectra, only a first order fit to baseline is removed, showing that CND and polar arc spectra suffered from strong baseline fluctuations. 
\label{nondetect}
\end{figure*}

Toward three positions, we do not detect either the \am\, (8,8) or (9,9) lines: the polar arc, M0.16-0.10, and M-0.32-0.19 (Figure \ref{nondetect}). Upper limits on the line intensities for these clouds are reported in Table \ref{results}. The polar arc is a higher-latitude filament extending at a 40\degr\, angle from ({\em l,b} = 0\degr,0.05\degr) to ({\em l,b} = 0.2\degr,0.25\degr) which is seen in CO and CS surveys of the Galactic center \citep{Bally88}. There is a strong velocity gradient along the arc, from 70 to 140 \kms, but it is not clear from its kinematics where this feature lies along the line of sight. The other two positions correspond to Galactic center clouds for which temperatures in excess of 500 K have been inferred by \cite{RF01} using mid-infrared observations of pure rotational lines of H$_2$.  We discuss these non-detections and their significance further in Section \ref{properties}. 

\subsection{Detection of Non-metastable Ammonia}
We also detected several low-excitation non-metastable \am\, lines toward the two strongest sources of metastable \am: M-0.02-0.07 and M-0.13-0.08. Non-metastable \am\, lines have previously been detected in the Galactic center cloud Sgr B2 \citep{Zuckerman71,Huttem93}, but only upper limits existed for M-0.02-0.07 and M-0.13-0.08 \citep{AB85}. The (2,1) and (3,2) lines were detected toward both clouds, and the (4,3) line was additionally detected toward M-0.13-0.08. The line parameters are listed in Table \ref{results}. Whereas M-0.02-0.07 has stronger metastable line emission, M-0.13-0.08 exhibits stronger non-metastable line emission. As we discuss in Section \ref{densities}, non-metastable line emission requires either a strong far-infrared radiation field or very high densities to populate these transitions, so it is possible that the non-metastable line emission originates in a gas component distinct from the metastable emission, since the same constraints do not apply to the metastable lines.
 
\subsection{Ammonia Temperatures}
\label{RT}

\subsubsection{Column Densities}
\label{cd}
To determine the rotational temperatures using Boltzmann statistics, we must first determine the level populations from each rotational transition. We can only measure temperatures for those clouds for which we detect at least two lines of either para ($K \ne 3n$, where $n$ is an integer) or ortho ($K = 3n$) \am, as ortho- and para-\am\, have different spin-alignment states, so transitions between para- and ortho-\am\, are forbidden, and the two behave as separate species. 

For lines observed in emission, assuming the emission to be optically thin, the beam-averaged column density can be calculated as:

	\begin{equation} N (J,K) = \frac{1.55\times10^{14}\, \mathrm{cm}^{-2}}{\nu}\frac{J(J+1)}{K^2}\int T_{\rm mb}\, \mathrm{dv}\,,
	\label{eqd}
	\end{equation}
	
\noindent as in \cite{Mauers03}. Here, the transition frequency $\nu$ is in GHz, and $\int T_{\rm mb}\,$dv is the integrated intensity of the line, in units of K \kms. 

For lines observed in absorption against the background continuum, to determine the column density requires
knowledge of the excitation temperature across the inversion doublet. As we do not know this quantity without radiative transfer modeling, which we do not undertake in this paper, we thus report the quantity $N/\mathrm{T}_{\rm ex}$, as follows:

	\begin{equation} \frac{N}{\mathrm{T}_{\rm ex}} = \frac{1.61\times10^{14}}{\nu} \frac{J(J+1)}{K^2} \tau \Delta v_{1/2}\, \mathrm{cm}^{-2} / K\,,
	\label{eqe}
	\end{equation}
	
\noindent as in \cite{Huttem95}. Here, $ \Delta v_{1/2}$ is the FWHM of the line in \kms, and $\tau$ is the line optical depth, which 
can be measured directly from the brightness temperature of the line relative to the continuum:

	\begin{equation} \tau = -ln \left(1 - \frac{|T_L|}{T_C} \right). 
	\label{eqf}
	\end{equation}

We also observed the (1,1) and (2,2) lines in M-0.02-0.07. These lower-excitation lines have prominent hyperfine structure made up of multiple satellite lines spaced a few tens of \kms\, from the main line. As the relative LTE line strengths of the satellite lines are a small fraction\footnotemark[1]\footnotetext[1]{The ratio of the intensities of the hyperfine satellites to the main line are 0.28 and 0.22 for the inner and outer pairs of satellite lines to the (1,1) line, and 0.06 for the both pairs of satellites to the (2,2) line} of the total line intensity, the hyperfine satellites are generally optically thin, and so the ratio of the satellite line strengths to the main line provides a good measure of the line opacity. However, for the wide linewidths typical of Galactic center clouds, these satellite lines are strongly blended with the main line, and the intrinsic line widths are uncertain. First, we fit to the hyperfine structure and estimate the opacities using the FIT routine in the CLASS software package\footnotemark[2] \footnotetext[2]{http://www.iram.fr/IRAMFR/GILDAS} by assuming that the (1,1) and (2,2) line widths are the same as that measured for the (3,3), (4,4) and (6,6) lines: 22.7 \kms (Table \ref{hyperfine}). The fits to these lines are shown in Figure \ref{hyperfig}. 

However, if the line width is not fixed, there is a wide range of equally good fits for widths from 17 to 30 \kms. We also report in Table \ref{hyperfine} the range of parameter space for which the fit is at least as good as the fixed-width fit. Comparably-good fits are found for opacities ranging from 0.1 to 1.3 for the (1,1) line and from 0.1 to 3.6 for the (2,2) line. For comparison, \cite{Huttem93b} find (1,1) and (2,2) opacities of $\sim5-5.7$ for M-0.13-0.08, which has comparable (and in fact slightly lower) hot \am\, column densities. This difference in opacity is likely because the linewidths in this cloud are much narrower (\citeauthor{Huttem93b} measure the widths of the (1,1) and (2,2) lines to be $\sim$10 \kms, or half of that which we measure for M-0.02-0.07).
 
We calculate the column density of the optically thick (1,1) and (2,2) lines in M-0.02-0.07 according to Equation 1 from Appendix B of \cite{Huttem93b}:

	\begin{equation} N (J,K) = 6.8\times10^{12}\, \mathrm{cm}^{-2}\frac{J(J+1)}{K^2}\int T_{ex}\, \tau\mathrm{dv}.
	\label{eqg}
	\end{equation}
	
For these two lines, the calculated column densities are the actual, rather than beam-averaged column densities. We convert these column densities to beam-averaged column densities by assuming that the emission from these low-excitation lines fills the beam, and the beam filling factor is equal to one. If the actual beam filling factor is smaller than one, then the column densities we report are upper limits on the true beam-averaged column densities. The calculated column densities for all of the detected lines are given in the last column of Tables \ref{results} and \ref{hyperfine}a. 

\subsubsection{Rotational Temperatures}
\label{rtemps}

\begin{figure*}
\includegraphics[scale=0.4]{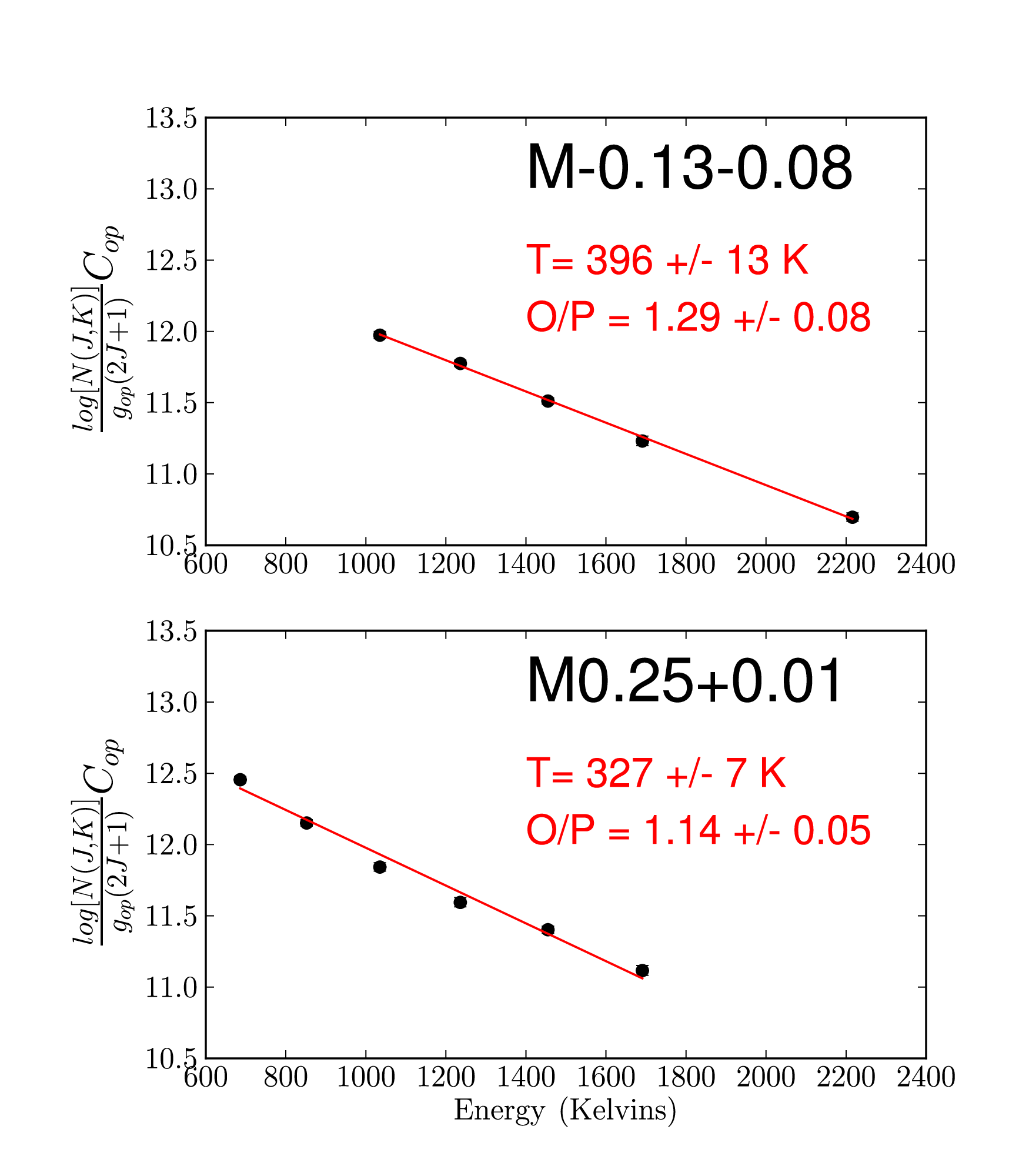}
\includegraphics[scale=0.45]{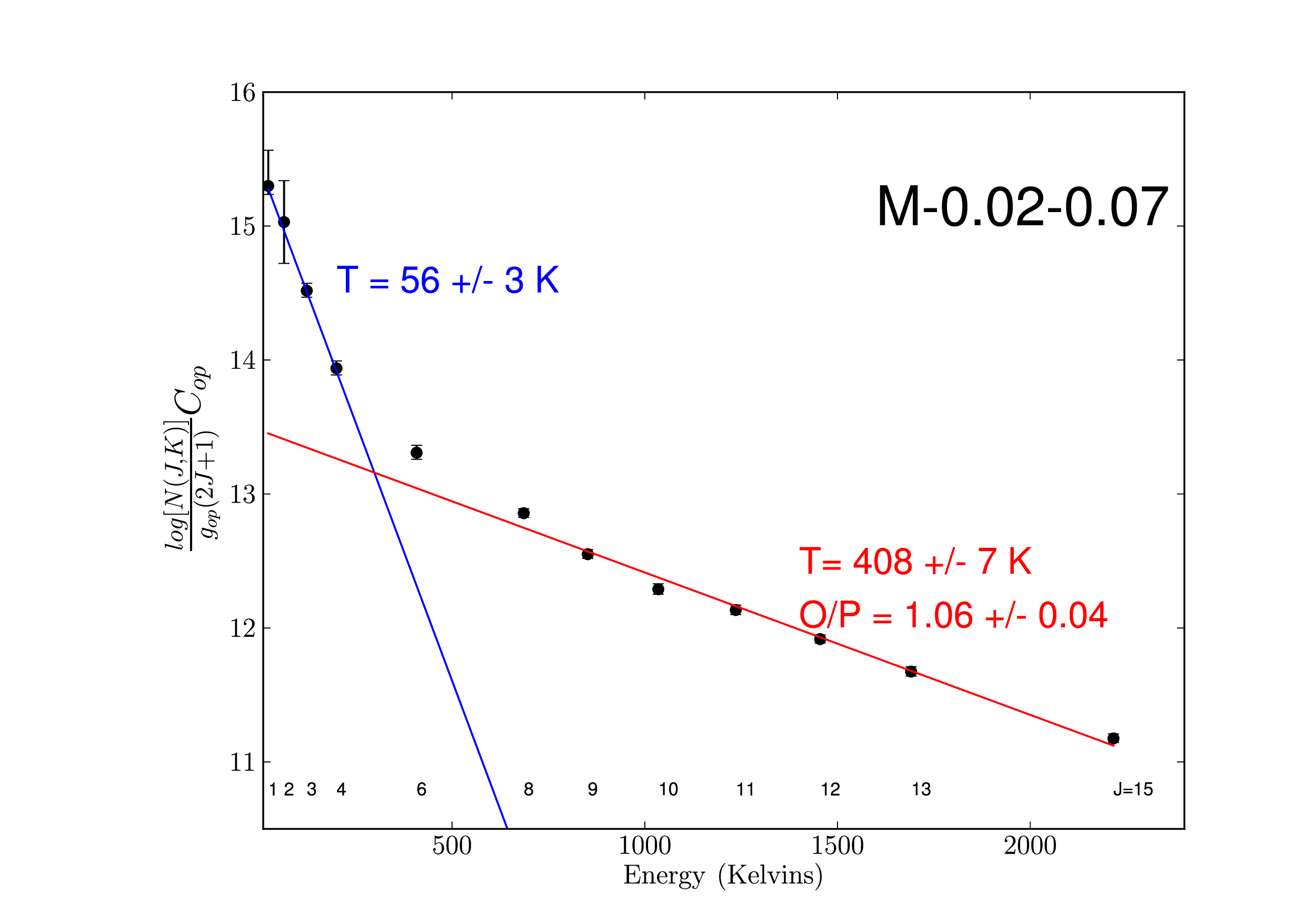}
\caption{ Rotational \am\, temperatures determined for three clouds. The temperatures are fit to the slope of the normalized \am\, column density (corrected for the level degeneracy) plotted against the energy of each transition in units of Kelvins. The plotted column densities are multiplied by an additional factor, $C_{op}$, which is equal to one for lines of ortho-\am, and for lines of para-\am\, is equal to the ortho-to-para ratio (O/P) calculated using the $J\geq 8$ lines.  For the $J \leq 6$ lines in M-0.02-0.07, we multiply by the $C_{op}$ determined for the higher-$J$ lines.    }
\label{temp_plot}
\end{figure*}

The rotational temperature of \am\, can be determined by plotting the logarithm of the normalized column density: $log_{10}[ N(J,K) / (g_{op} (2J + 1)) ]$, where the statistical weight factor, $g_{op}$, is 2 for ortho-\am\, ($K =$ 9, 12, 15), and 1 for para-\am\, $(K = 8, 10, 11,$ ...), against the upper level energy of each rotational state (Figure \ref{temp_plot}). From the Boltzmann equation, the rotational temperature is then related to the slope $m$ as T$_{rot}$= -$log_{10}$(e)/$m$. As described previously, the rotational temperature is close to but less than the kinetic temperature, T$_{\rm K}$\citep{Morris73, Danby88}. The amount by which the rotational temperature underestimates the true kinetic temperature is lessened when the upper level energy of the highest transition used to measure the temperature is equal to or greater than the kinetic temperature of the gas being measured. However, rotational temperatures determined using transitions between lower $(J,K)$ levels will significantly underestimate the true kinetic temperature of a hot gas component. When the logarithm of the level populations implied by observations is plotted against the upper level energy (a Boltzmann diagram, e.g., Figure \ref{temp_plot}), one can see this effect manifested as a slightly upward-concave curve instead of a straight line expected for a constant T$_{\mathrm K}$. This is especially apparent for the lowest-excitation metastable lines of \am, \citep[c.f.,][]{Huttem93b}.

As we have both ortho- and para- species of \am, we perform a simultaneous fit for both the slope of the line and the ortho/para ratio, or the multiplicative factor between the two species required for the best fit lines to have the same intercept.  Since, as previously discussed, the measured rotational temperature is a lower limit to the actual kinetic temperature, and the more highly-excited lines yield a better limit for a high kinetic temperature than lower lines, we also determine the rotational temperatures between the two highest para-lines ($K=11,13$), and between the two highest ortho-lines ($K=12,15$). %This analysis assumes that all of the observed ammonia lines arise from the same volume within the cloud, and furthermore, that there are no systematic differences in the distribution of \am\, within each cloud due to chemistry. 

Temperatures for the three clouds for which we observe \am\, lines up to $(J,K)$ = (15,15) are reported in Table \ref{temps}. We find best-fit rotational temperatures using all measured lines that range from 340 K for M0.25+0.01, up to 410 K for M-0.02-0.07. The rotational temperatures calculated between the most highly-excited transitions, T$_{(15,15)-(12,12)}$ and T$_{(13,13)-(11,11)}$, are systematically higher, ranging from 410 K to 450 K. 

For M-0.02-0.07 we also mapped the (1,1) through (4,4) and (6,6) lines. Using the spectra extracted from these maps at the same position as we observed the (8,8) through (15,15) lines in this cloud we can also derive rotational temperatures from the lower-excitation lines of \am. We derive rotational temperatures of $\sim$ 50 K for the para-\am\, lines, and $\sim$ 100 K for the ortho-\am\, lines (Figure \ref{temp_plot}). Using the results from the radiative transfer models of \cite{Ott2011}, these rotational temperatures correspond to kinetic temperatures of $\sim$ 70 K (from the (2,2)-(4,4) rotational temperature), and $\sim$ 150 K (from the (3,3)-(6,6) temperature). However, as previously noted, the column densities of the (1,1) and (2,2) lines are quite uncertain, due to difficulty in fitting the hyperfine structure. If the column densities of these two lines are underestimated, then there may also be a cooler gas component present, more consistent with the $\sim25$ K component detected by \cite{Huttem93b} toward a large sample of CMZ clouds.  
 
\subsubsection{Ortho/Para ratio}
\label{o_p}

As part of our temperature fits using the highly-excited \am\, transitions observed toward M0.25+0.01, M-0.02-0.07, and M-0.13-0.08, we also derive the ortho/para ratio for each cloud. Because radiative and gas-phase collisional transitions are not allowed between ortho and para states, the ortho/para ratio is representative of the conditions (either the gas temperature, or if the \am\, is formed via grain surface reactions, the grain temperature) at the time of formation of the molecules. We find ortho/para ratios ranging from 1.0  to 1.3, which indicate that the molecules formed in ``high"-temperature (T $\gtrsim$40 K) conditions \citep{Tak02}, that is, temperatures that are consistent with the inferred temperatures of Galactic center molecular clouds.  

Finally, for the majority of sources presented here, we only observed the \am\, (8,8) and (9,9) lines. As the (8,8) line is a transition of para-\am, and the (9,9) a transition of ortho-\am, it is necessary  to assume an ortho/para ratio in order to determine a temperature using these lines. Adopting an ortho/para ratio of 1.1 (the average value from the fits to M0.25+0.01, M-0.02-0.07, and M-0.13-0.08 reported in Table \ref{temps}), we find rotational temperatures for all but one of these sources (G1.6+0.025, described below) which range from $\sim$200 to 320 K. For comparison, T$_{8,8-9,9}$ for M-0.02-0.07 calculated using the same ortho/para ratio is only $\sim$ 220 K. %This suggests that almost of the clouds in which we detect these highly-excited transitions of \am\, have a temperature component which is similarly hot (or perhaps hotter than) the $\sim$ 400 K component we detect in M-0.02-0.07. For Sgr B2-M, this is indeed the case, as \cite{Wilson06} have measured a hot gas component with T $\sim$ 600 K in this cloud. 

For three sources, G1.6+0.025, M0.34+0.06, and G0.18-0.04,  if we use the marginal detection or upper limit on the (8,8) line strength, then an assumed ortho/para ratio of 1.1 does not yield a meaningful temperature. For these clouds, it appears that the ortho/para ratio may be slightly anomalous. 
If we assume a temperature of 200 K (as measured from lines of CH$_3$OH in G1.6+0.025 by \cite{Menten09} ), then we find ortho/para ratios of 2.6, 2.0, and 2.4 in G1.6+0.025, M0.34+0.06, and G0.18-0.04, respectively. If we assume a temperature of 400 K, then the ortho/para ratios would be correspondingly reduced: 1.7, 1.3, and 1.6. 
These higher ratios suggest that the \am\, in these clouds may have formed in relatively low-temperature conditions: $<20$ K \citep{Tak02},

\section{Discussion}

From the observed sample of 17 pointings toward CMZ molecular clouds, we detect emission from  the highly-excited (9,9) inversion transition of \am\, in thirteen locations. For three of the strongest \am\, sources, we further observe \am\, lines up to (15,15) and measure rotational temperatures which, as discussed previously, are lower limits to the actual kinetic temperature of the gas. Below, we compare the temperatures we find to previous \am\, temperature measurements, and discuss the distribution of these clouds in the CMZ, he nature of this hot gas component, and its potential heating sources. Properties of individual sources in which we detect hot \am\, are described in the Appendix %\ref{individual}.

\subsection{Temperature ranges. }

We find that the highest-excitation \am\, lines we observe have rotational temperatures of 350 - 450 K gas in each of the three clouds for which we were able to determine temperatures. These temperatures are rotational temperatures, and thus a lower limit on the kinetic temperature, which must be greater than or equal to these values.  As described in Section \ref{rtemps}, these temperatures are derived from both ortho- and para-\am\, lines with a simultaneous fit for temperature and the ortho/para ratio. The temperatures derived from just the ortho-\am\, lines appear slightly higher than those derived from just the para-\am\, lines, however this difference is generally within our measurement uncertainty, so it is not clear that this is a real effect. The observed ortho lines also have higher level energies, so may be more sensitive to presence of hotter gas, yielding rotational temperatures which are closer to the true kinetic temperature of the gas. 

Previous temperature measurements in CMZ molecular clouds using metastable \am\, inversion lines from (1,1) to (6,6) indicated both a 25-50 K temperature component , and a 200 K temperature component\citep{Gusten85,Huttem93b}. As the lines used in these temperature analyses are lower-excitation lines of \am, they are less sensitive to the presence of hot gas, and likely do trace intrinsically cooler gas in the cloud. However, the 200 K temperature component reported by \cite{Huttem93b} is actually a rotational temperature, and thus a lower limit to the true kinetic temperature of the gas. Using the results of large-velocity-gradient modeling of \am\, excitation \citep{Ott2005,Ott2011} which relate observed rotational temperatures to kinetic temperatures, we find that 2/3 of the \cite{Huttem93b} rotational temperatures derived from the (4,4)-(5,5) \am\, lines are consistent with a kinetic temperature of $>$ 300 K. Observations of the (5,5) and (7,7) \am\, lines by \cite{Mauers86} in the same three clouds for which we measure temperatures also yielded higher rotational temperature estimates of $\sim$300 K. 
This suggests that the hot gas (T $\geq$ 350-450 K) we detect with highly-excited lines of \am\, could be the same as the warm gas component previously detected by \cite{Huttem93b} and \cite{Mauers86}, and consistent with a model of just two temperature components in GC clouds: a cooler component of 25-50 K, and a hot component of $>$ 300 K. We test this idea in Section \ref{distribution} using our observations of M-0.02-0.07, which include the (1,1) through (6,6) lines. 

\subsubsection{Caveats. }

The temperatures we derive are based on several assumptions. First, we assume that the observed emission is roughly uniform over an area larger than the beam sizes of these observations. If the emission arises from a source that is more compact than the smallest beam (20$''$), then the observed brightness temperatures would need to be corrected for the different beam size at each frequency. For M-0.02-0.07, in the extreme case that the \am\, emission originated in a point source, we would determine a rotational temperature of $\sim$370 K, instead of $\sim$450 K. However, we find it extremely unlikely that the highly-excited \am\, arises in compact sources. As we show below in Section \ref{distribution}, in M-0.02-0.07 we find that the majority of the \am\, (9,9) flux arises on scales larger than tens of arcseconds. We also assume that the excitation temperature T$_{ex}$ is the same for all of the \am\, lines measured, which may not be valid over the wide range of level energies these transitions cover. \cite{Wilson06} estimate for Sgr B2 that the variation in T$_{ex}$ is roughly a factor of two: decreasing from 6 K for \am\, (1,1) to 3 K for \am\, (18,18). %First, we assume that all of the observed lines arise from the same volume within the cloud.  We also assume that chemistry does not have a significant effect on the distribution of ammonia within individual clouds. 

The upper level energies of \am\, assumed for these calculations, from the JPL Submillimeter, Millimeter, and Microwave Spectral Line Catalog \citep[Table \ref{ammonia};][]{JPL}, are systematically larger than upper level energies taken from the Spectral Line Atlas of Interstellar Molecules (SLAIM) \citep[Available at http://www.splatalogue.net, F. J. Lovas, private communication;][]{Remijan2007}. The difference in level energies increases from 0.1\%, for the (8,8) transition to 1.6\%, for the (15,15) transition. Using the lower SLAIM values would result in temperatures 10-20 K lower than we report, a difference which is slightly larger than our estimated uncertainty. 
 
\subsection {The nature of the 400 K gas component }

\subsubsection{Distribution of \am\, (9,9) in the CMZ}

The positions where we detect \am\, (9,9) are distributed throughout the central 3 degrees of the Galaxy. We detect this line in G1.6-0.25, which is 240 pc in projection to the east of Sgr A* \citep[assuming R$_0$ = 8.4 kpc; ][]{Ghez08, Gillessen09}, as well as in Sgr C (R = 80 pc to the west). We also detect the (9,9) line very strongly in M0.25-0.01, which in the infrared continuum appears strongly in absorption against the extended background emission \citep{Longmore}, suggesting that it is located at the front edge of the CMZ. The detection of the (9,9) line toward all of these positions implies that hot \am\, is present in GC molecular clouds over a broad range of galactocentric radii. However, there are too few clouds in our sample for which we determine temperatures to be able to identify any correlations in the distribution or temperature of the hot gas with respect to location in the CMZ. We can say however that the presence of hot gas in a cloud does not require the cloud to be actively forming stars, as M0.25+0.01 is apparently almost entirely quiescent, having no embedded warm IR sources or compact \hii regions.

\begin{figure*}
\hspace{1cm}
\includegraphics[scale=0.4]{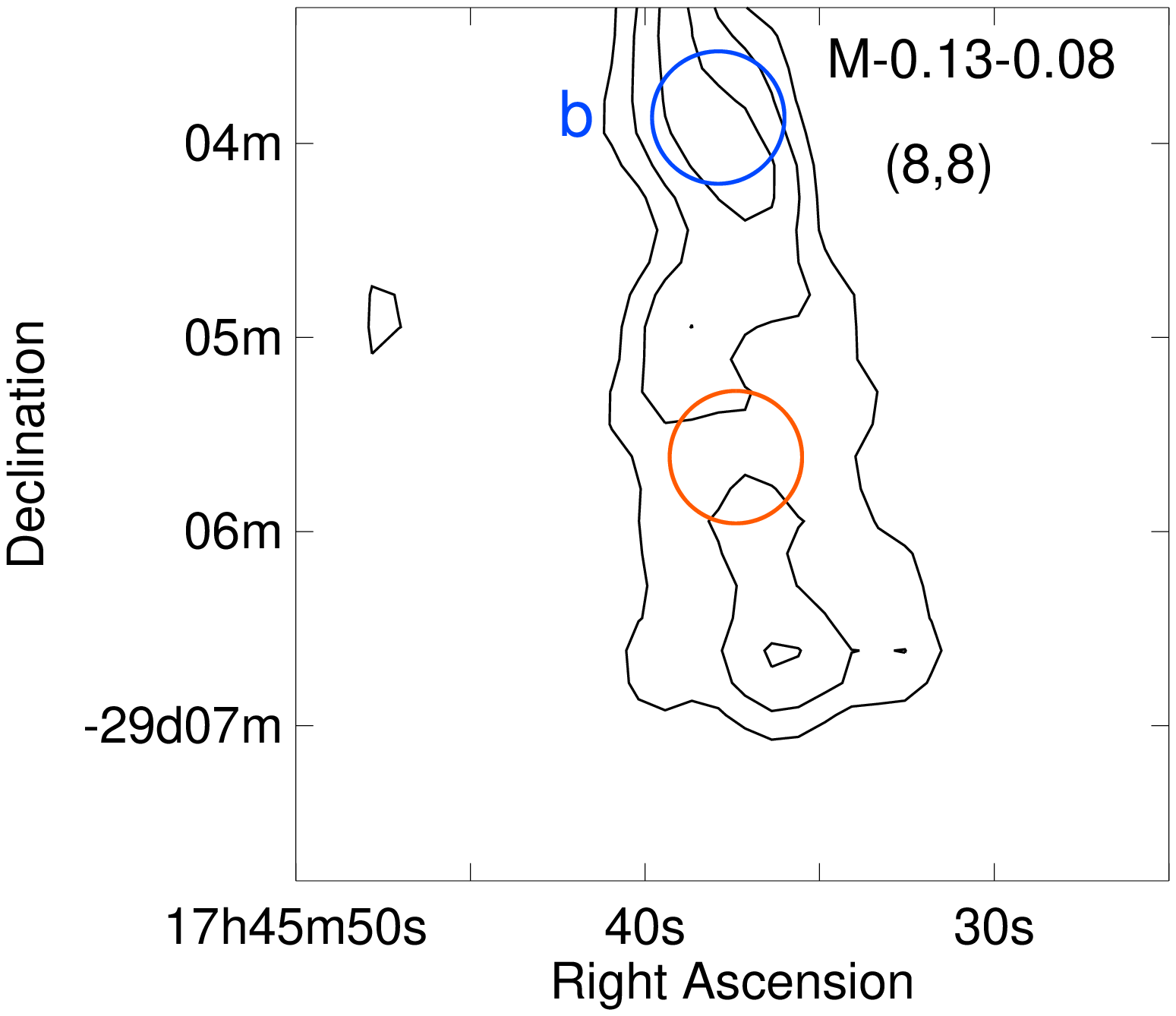}
\includegraphics[scale=0.4]{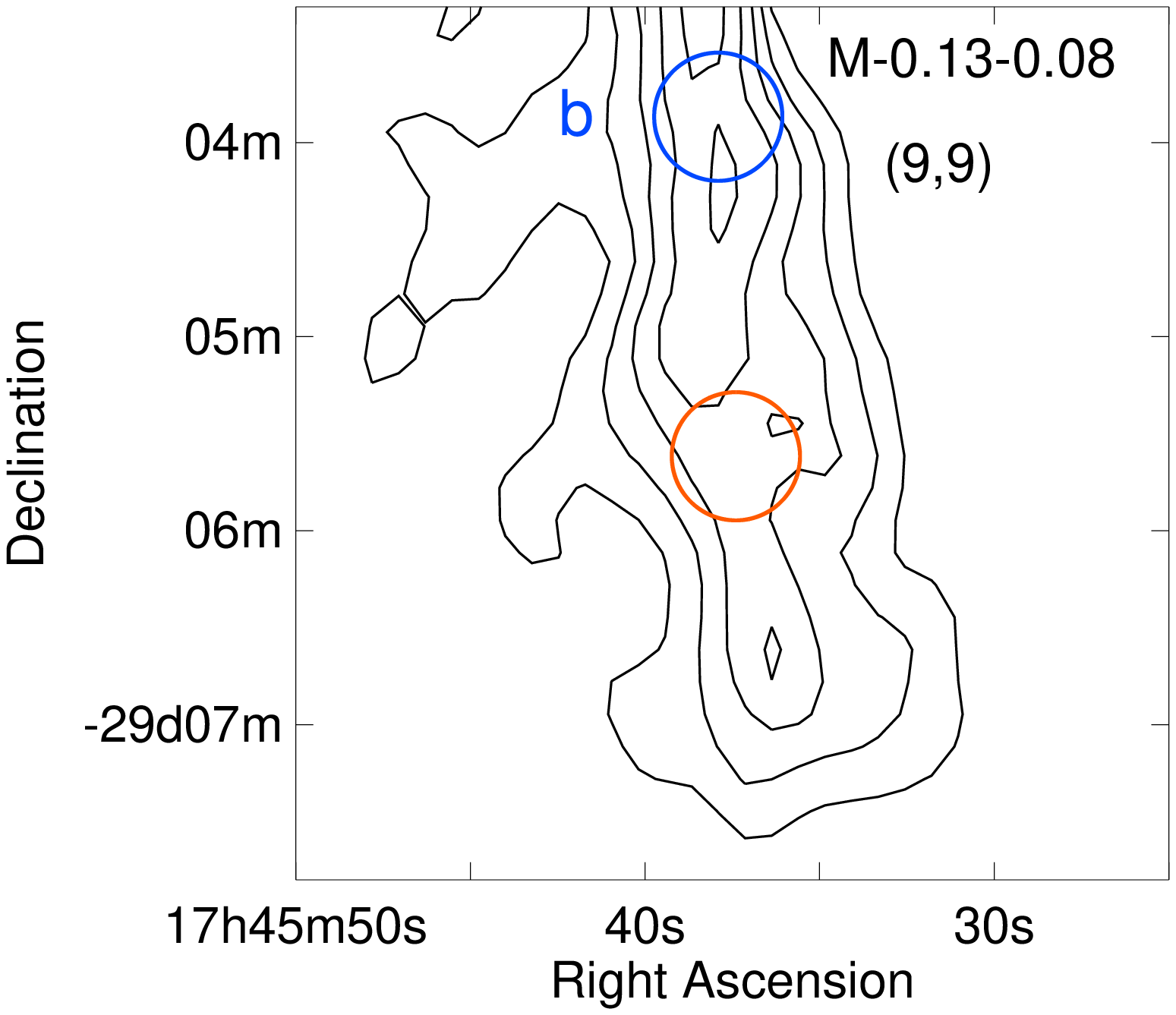}
\vspace{1cm}
\caption{ Maps of \am\, (8,8) and (9,9) integrated main-beam brightness temperature in M-0.13-0.08. The contour levels are linearly spaced by 1.5 K \kms\,, with the lowest contour being 4.0 K \kms. The peak integrated brightness temperatures  in the (8,8) and (9,9) maps are 7.8 K \kms\, and 10.8 K \kms, respectively. The noise levels in the images are 1.0 K \kms\, and 0.7 K \kms, respectively. }
\label{map_20}
\end{figure*}

\begin{figure*}
\hspace{1cm}
\includegraphics[scale=0.8]{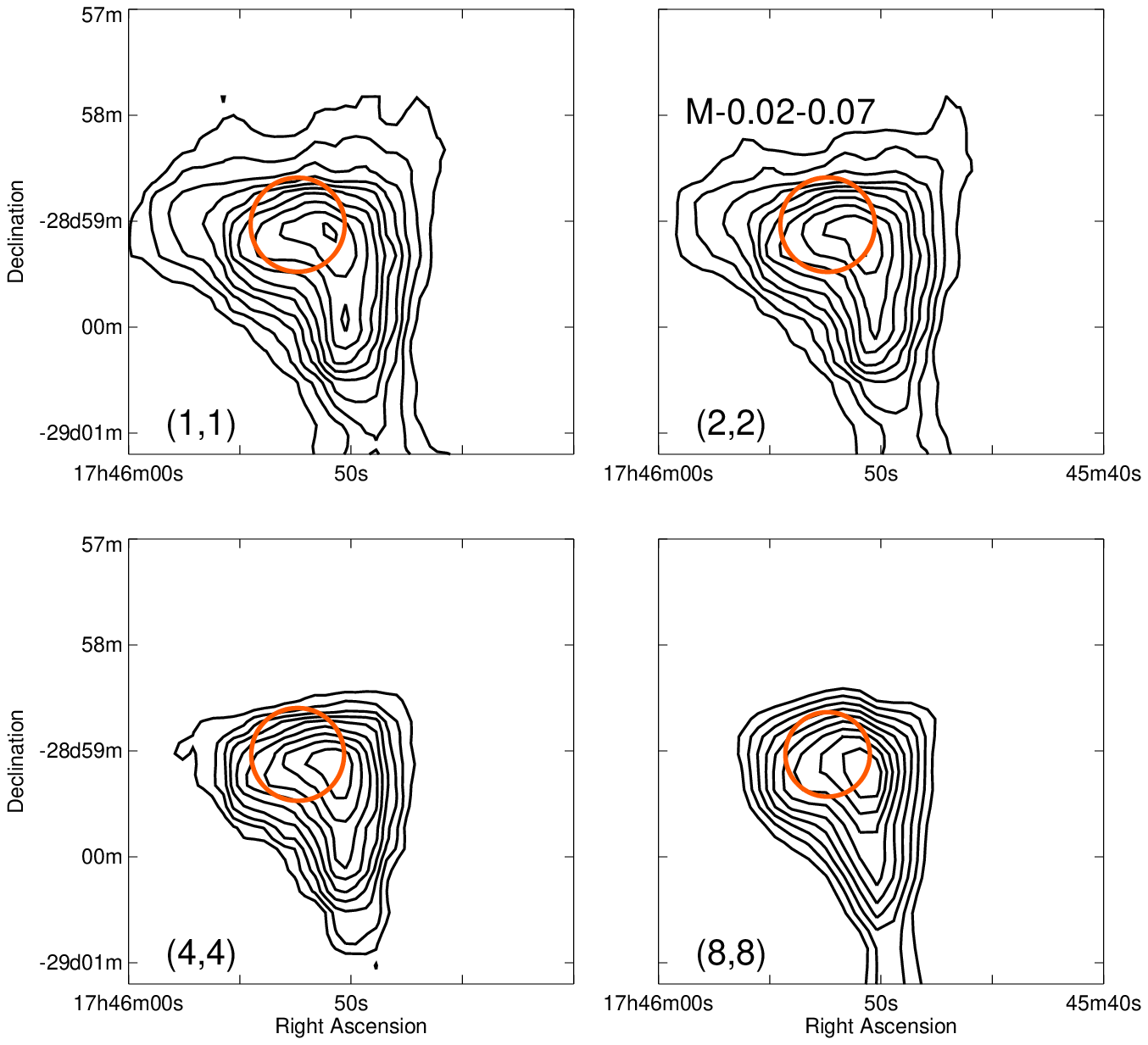}
\vspace{1cm}
\caption{Maps of integrated main-beam brightness temperature for the para-\am\, lines in M-0.02-0.07.  The contours are linearly spaced from the peak integrated brightness temperature (290, 236, 106, and 16.4 K \kms, for the  (1,1), (2,2), (4,4) and (8,8) lines, respectively) to 1/10 of this value. The noise levels in each map are 5, 5, 4, and 0.8 K \kms, respectively. }
\label{map_para}
\end{figure*}

\begin{figure*}
\hspace{1cm}
\includegraphics[scale=0.8]{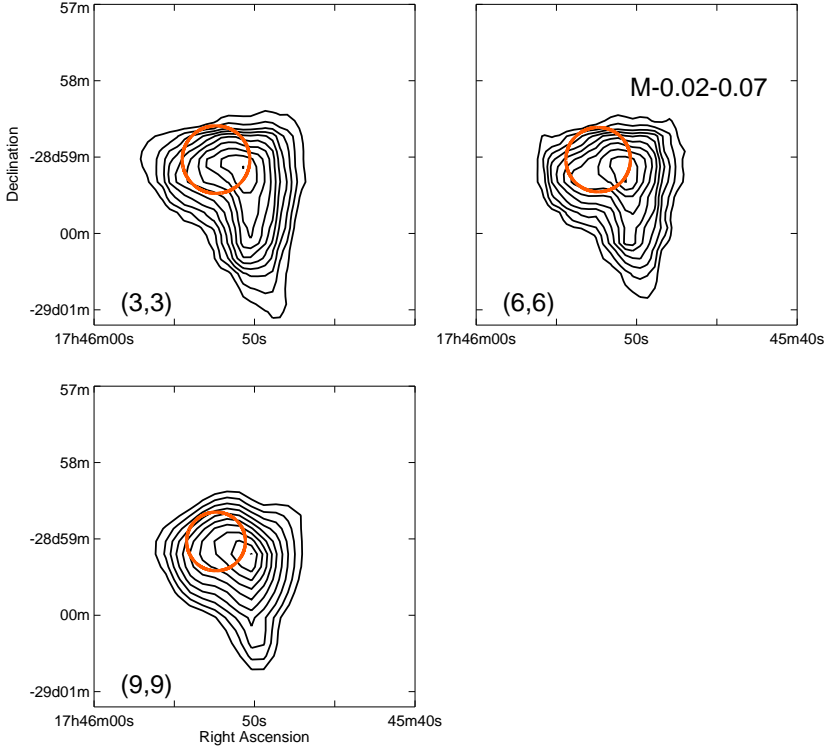}
\vspace{1cm}
\caption{Maps of integrated main-beam brightness temperature for the ortho- \am\, lines in M-0.02-0.07.  The contours are linearly spaced from the peak integrated brightness temperature (608, 91, and 25.1 K \kms\, for the (3,3),(6,6), and (9,9) lines, respectively) to 1/10 of this value.  The noise levels in each map are 6, 3, and 1.0 K \kms, respectively. }
\label{map_ortho}
\end{figure*}

Our maps of highly-excited \am\, in two clouds in the central 10 pc (M-0.13-0.08 and M-0.02-0.07, Figures \ref{map_20},  \ref{map_para}, and \ref{map_ortho} ) further show that the hot \am, as traced by the (8,8) and (9,9) lines, is truly concentrated in the dense centers of the GCMCs. The distribution of emission from these higher-excitation lines is almost identical to that of the lower-excitation (1,1) through (6,6) lines, which should trace, cool, dense gas. The $1.5'-2'$ extended sizes of the hot \am\, clouds, and their morphologies, are additionally very similar to those of the submillimeter and millimeter continuum emission from cool dust in the cores of these clouds. We also find that the hot \am\, emission is morphologically and kinematically similar to emission from the dense-gas-tracing HC$_3$N 3-2 line, which we also observe in our GBT data.

As the hot gas component traced by highly excited lines of \am\, appears to be associated with the dense cores of molecular clouds, it is qualitatively different than another warm gas component (T= 200-350 K, n = 50-100 cm$^{-3}$) which has been detected in the CMZ from the absorption of H$_3^+$ along multiple lines of sight \citep{Oka05, Goto08, Goto11}. In addition to being cooler and more tenuous, the gas component traced by H$_3^+$ is suggested to widespread throughout the entire CMZ \citep{Goto08}, in contrast to the concentrated distribution of hot \am\, that we observe toward the cores of GCMCs. 

\subsubsection{\am\, in M-0.02-0.07}
\label{distribution}

\begin{figure*}

	\includegraphics[scale=0.3]{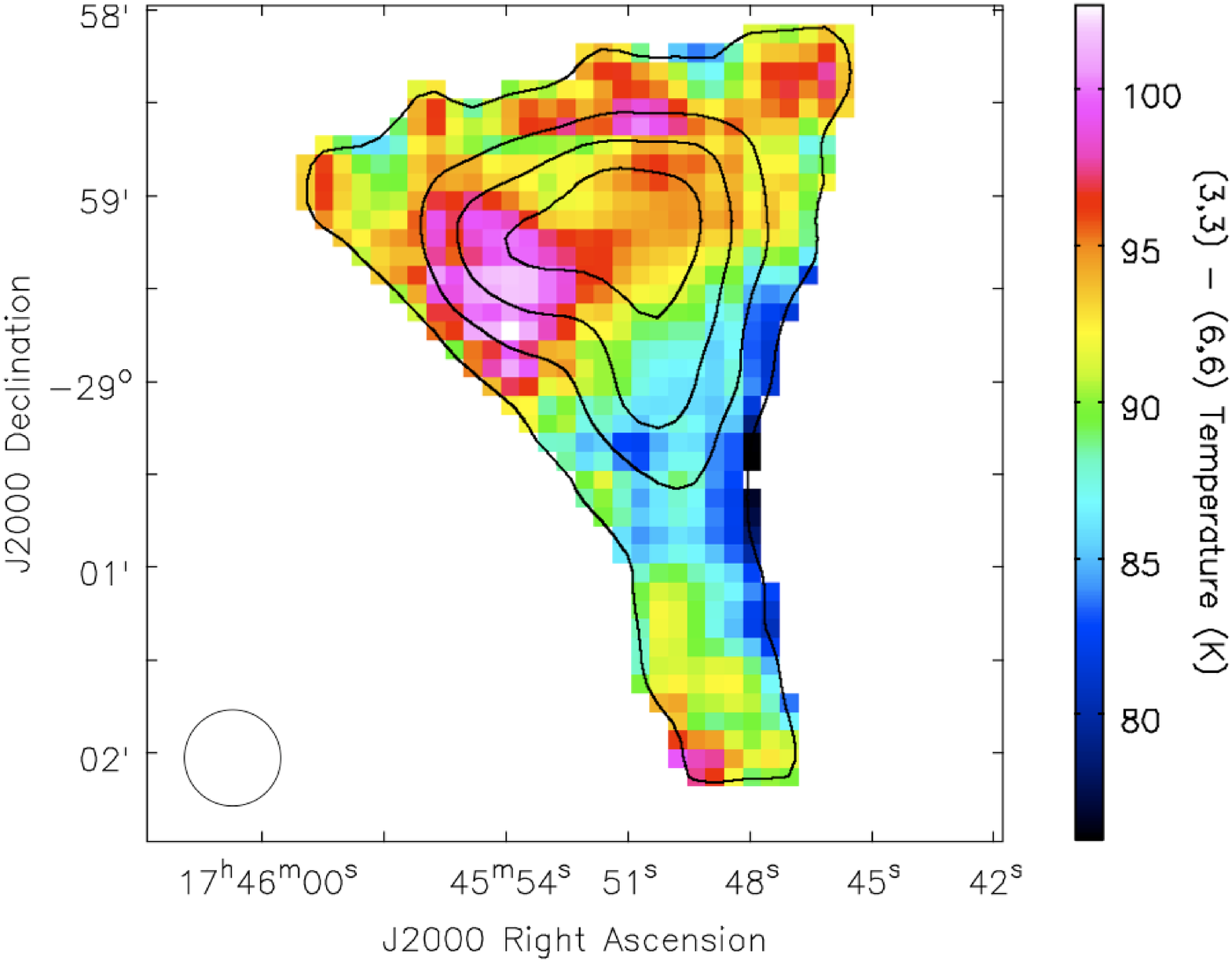}

	\includegraphics[scale=0.3]{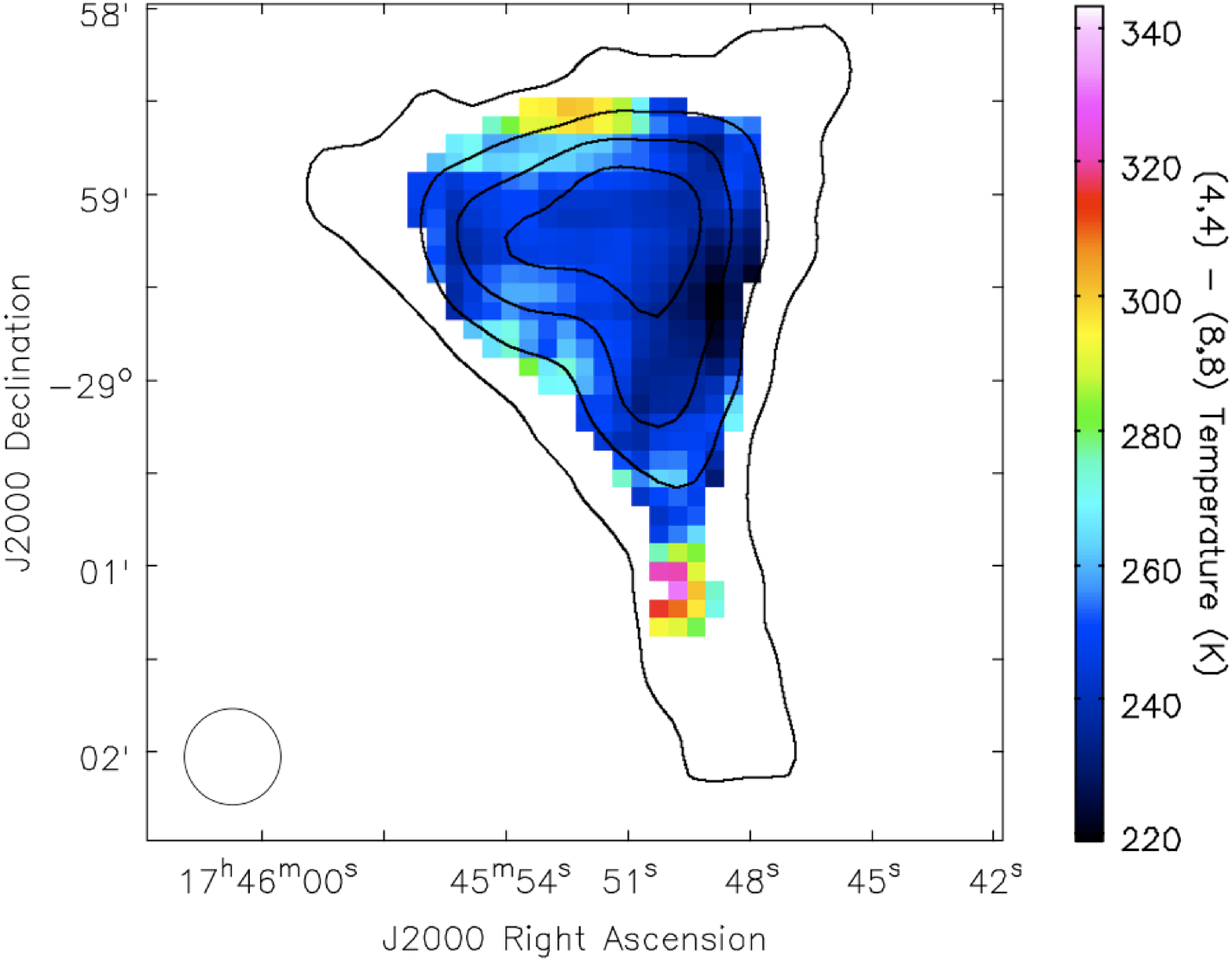}

	\includegraphics[scale=0.3]{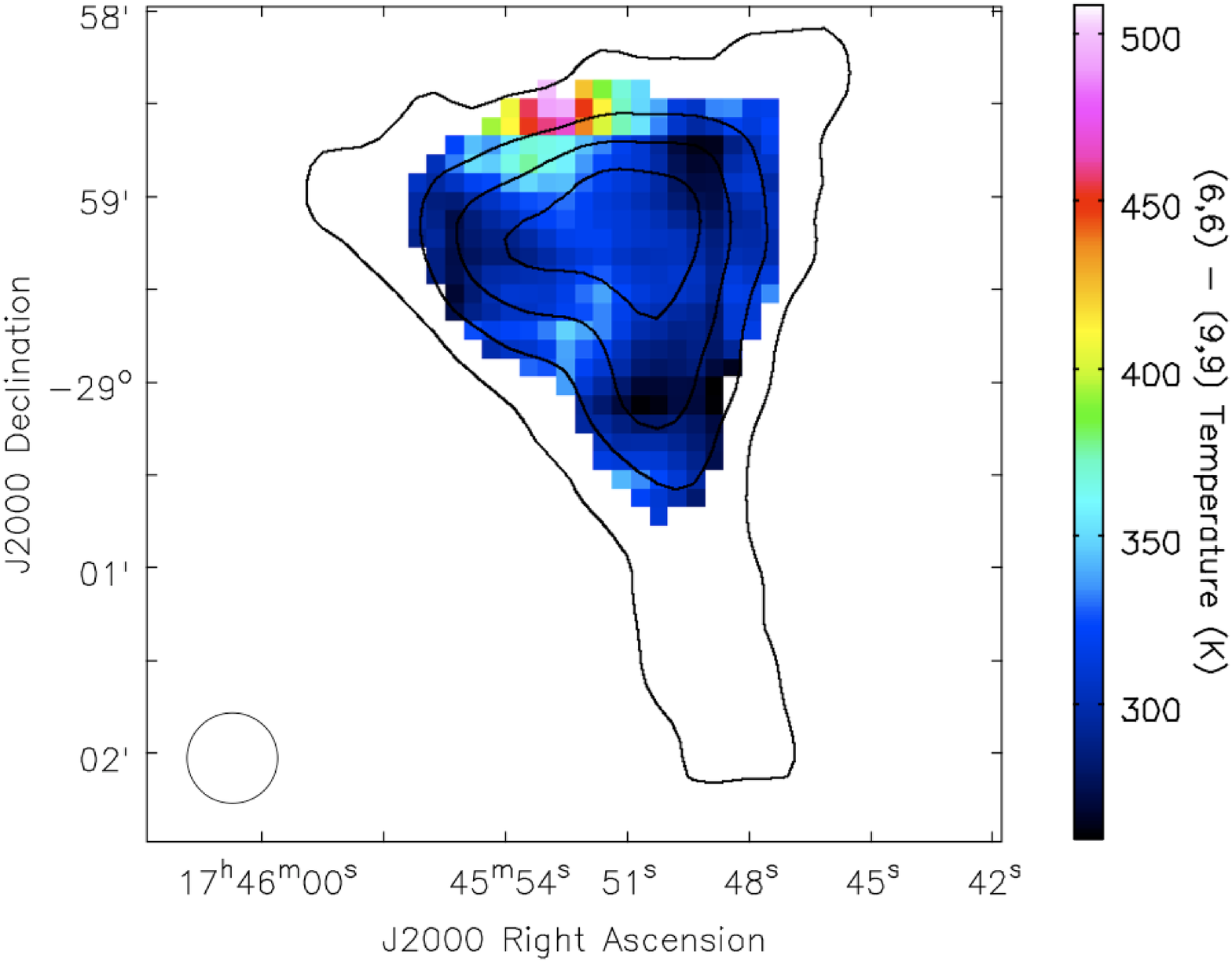}

\caption{Maps of the rotational temperature in M-0.02-0.07 derived from the \am\, (3,3) and (6,6) lines (Top), the  \am\, (4,4) and (8,8) lines (Middle), and the  \am\, (6,6) and (9,9) lines (Bottom). Contours of emission from the \am\, (6,6) line are overplotted in black. As the rotational temperature is a lower limit to the kinetic temperature of the cloud, the rotational temperatures measured with these relatively low-excitation lines underestimate the temperature of the hottest gas measured in this cloud using lines above (9,9). The apparent peak in temperature in the (4,4)-(8,8) temperature map at the southern edge of the cloud lies on the edge of the (8,8) map, and so may not be a real feature.}	
\label{temp_all}
\end{figure*}

We observe the largest number of \am\, transitions toward M-0.02-0.07, allowing us to probe in greater detail the properties of the hot \am\, component in this cloud. 
In particular, we have mapped emission from the (1,1), (2,2), (3,3), (4,4), (6,6), (8,8) and (9,9) lines, allowing us to derive maps of the rotational temperature for selected transitions. We avoid the (1,1) and (2,2) lines, due to their uncertain opacity, and present maps of the rotational temperature between the (3,3) and (6,6) lines, the (4,4) and (8,8) lines, and the (6,6) and (9,9) lines (Figure \ref{temp_all}). In order to create these maps, the higher-excitation line was convolved to the resolution of the lower-excitation line, and the maps were interpolated to have the same pixelization. Because the highest-excitation line used in these maps is the (9,9) line, the rotational temperatures shown in these maps are lower than the 400 K temperature component we derive for this cloud using transitions above (9,9). However, the average rotational temperatures do rise with the level of excitation of the two lines being used to construct the map.

All of the temperature maps show two main features. First, the variation in temperature across the cloud in all of the maps is relatively small: generally less than 50 K over the majority of the cloud. Secondly, as is especially apparent in the (3,3)-(6,6) temperature map, the western edge of the cloud is generally cooler than the eastern extension of the cloud. The western edge is the location of the ridge of compressed gas which is believed to have been swept up by the expansion of the adjoining Sgr A East supernova remnant \citep{Serabyn92}. All three maps also show some evidence for higher temperatures near the southeast edge of the cloud. One possible source of these higher temperatures could be the four compact \hii regions which are embedded in this part of the cloud \citep{Goss85,YZ10,Mills11}. 
The temperature maps using the (8,8) and (9,9) lines also show an additional, prominent feature: an apparent peak in the temperature on the northern edge of the cloud. This peak is not associated with any previously-identified feature of the cloud or its environment. We discuss potential heating mechanisms for this cloud in more detail in Section \ref{heating}

\begin{figure*}
\hspace{1cm}
\includegraphics[scale=0.9]{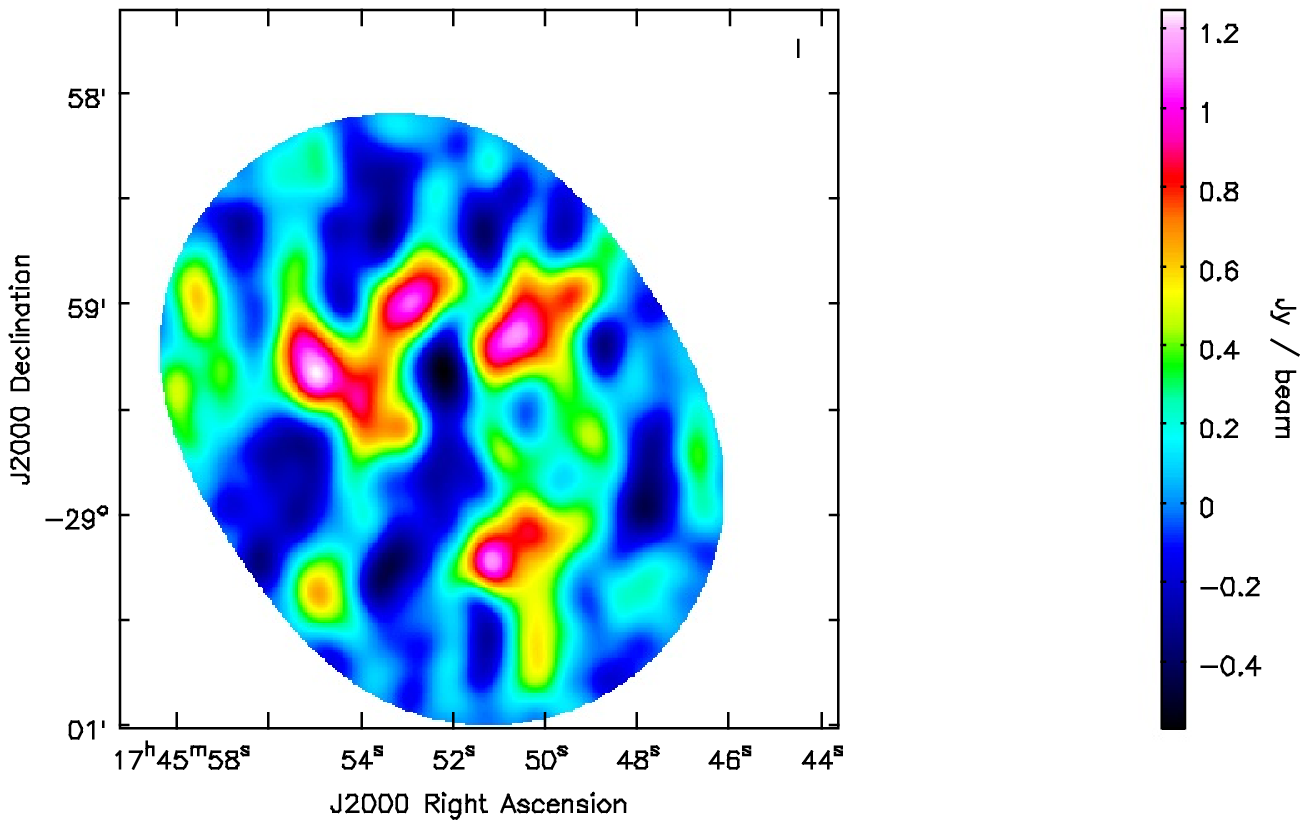}
\vspace{1cm}
\caption{Map of resolved \am\, (9,9) emission in M-0.02-0.07 from the VLA, convolved to a $10''\times10''$ resolution. The RMS noise in the image is 0.25 Jy beam$^{-1}$. }
\label{map_vla}
\end{figure*}

Although we are able to construct temperature maps of M-0.02-0.07, the resolution of our GBT data ($\sim 25''$ ) is insufficient to check whether the hot \am\, emission is truly uniform over the $1'-2'$ dense centers of GCMCs, or whether it originates in unresolved compact structures within the clouds. To investigate this, we also made VLA observations of the (9,9) line in M-0.02-0.07 to measure the fraction of emission resolved out on angular scales approaching that defined by the shortest VLA baseline used for these observations ($\sim60''$).  Figure \ref{map_vla} shows the resulting VLA image of \am\, (9,9) in M-0.02-0.07, smoothed to 10$''$ resolution. Although the emission appears clumpy, the total integrated flux detected with the VLA is 26 Jy \kms, compared to a total integrated flux of 660 Jy \kms\, derived from our GBT observations. This means that 96\% of the emission is resolved out, originating from structures on tens of arcsecond scales, and so any apparent clumps are likely just minor peaks in this extended emission. We note that a significantly larger fraction of \am\, (9,9) emission is resolved out by the VLA than was the case with \am\, (3,3) ($\nu = 23.870$ GHz) in the similar M-0.13-0.08 cloud, 10 pc away in projection from M-0.02-0.07 \citep{AHB85}. In M-0.13-0.08, \citeauthor{AHB85} found that 75\% of the \am\, (3,3) emission detected with a single-dish telescope was resolved out by the VLA. The largest recoverable angular scale should be $\sim$ 15\% smaller at 27.5 GHz than at 23.9 GHz, and so, were the emission from the (3,3) and (9,9) lines on the same size scale, one would expect only slightly more emission to be resolved out in our higher-frequency observations. The much larger fraction of flux which is resolved out for the (9,9) line suggests that this higher-excitation line originates in a more extended component than the lower-excitation (3,3) line. %However, as the \citeauthor{AHB85} observations of \am\, (3,3) were obtained with the old VLA receivers, this difference in the amount of flux which is resolved out could also be due in part to the lower sensitivity of older observations to faint structure. 

Although this would appear to favor an extended envelope as the origin of the highly-excited \am\, emission, we also consider that our maps of \am\, (8,8) and (9,9) emission in M-0.02-0.07 and M-0.13-0.08 do show the \am\, emission to be somewhat concentrated toward the cloud centers. This suggests that most of this hot gas is found in the regions of highest gas column density, and is potentially co-extensive with the highest-density gas and cool dust in these clouds. We suggest that this is more consistent with the highly-excited \am\, being distributed throughout the cloud, perhaps in a hot inter-clump medium, as opposed to being in a larger-scale envelope around the cloud's exterior. If in fact the hot \am\, originates in a relatively uniform network of unresolved shocks, this could explain both the large fraction of emission which is resolved out by an interferometer, and the apparent concentration of the hot \am\, emission toward the column density peaks of these clouds. 

In M-0.02-0.07, for which we detected 14 transitions of \am, we can determine the fraction of the measured \am\, column which arises from the most highly-excited transitions. To determine the total column density, we sum the contributions from all of the metastable levels up to $J$=15. We interpolate to estimate the column density for the $J$=5,7 and 14 levels, as we did not observe these lines. We also extrapolate to estimate the column density of the $J$=0 level using the column density of the $J$=1 level and the temperature determined from the (1,1), (2,2) and (4,4) lines ($\sim50$ K). In this way, we find that the total column density of \am\, in M-0.02-0.07 is $\sim$2.1$\times10^{16}$ cm$^{-2}$.  If we assume that all of the column density above $J$=10 originates in a hot component, which is at a temperature of 400 K, then we can calculate the fraction of the column density in lower lines which is also due to this hot component. In this way, we calculate that 12\% of the total measured \am\, column in M-0.02-0.07 arises in a T$\geq$400 K component (see Figure \ref{temp_plot}). 

However, the $J$=1 and $J$=2 column densities, which dominate the total column density, are also the most uncertain, due to difficulties in fitting the hyperfine structure of the (1,1) and (2,2) lines (See section \ref{cd}). We can, however, determine a lower limit on the fraction of the \am\, column which is hot by using the upper limits on the $J$=1 and $J$=2 column densities in the case that they are maximally opaque. Using these higher values (and further using a minimum temperature of 25 K to extrapolate to the column density of the $J$=0 level) we find an upper limit on the total column density of $\sim$2.9$\times10^{16}$ cm$^{-2}$.  The minimum fraction of the total \am\, column density in M-0.02-0.07  which arises from a T $\geq$ 400 K component is then 7\% (this fraction could also be higher if the beam filling factor of the (1,1) and (2,2) lines is less than 1, see discussion in Section \ref{cd}).  Note that this is only the fraction of \am\, which arises from this hot component; for example, if the \am\, abundance relative to H$_2$ is enhanced in the hot component, the fraction of $H_2$ which is hot could be less than 7\%. 

Finally, in Figure \ref{temp_plot} it is clear that there is a significant portion of the observed (6,6) and (8,8) column density (40\% and 25\%, respectively) which cannot be attributed either to a 50 K or 400 K temperature component. This excess column density indicates that two temperature components provide only an incomplete fit to the full range of temperatures in this cloud, which either requires an intermediate temperature component (like the $\sim$200-300 K component found by \cite{Huttem93b} or, more likely, a continuum of temperatures in order to properly fit all of the observed \am\, line intensities.

\subsection{Properties of clouds where \am\, (9,9) is not detected} 
\label{properties}
The three positions for which we do not detect emission from either \am\, (8,8) or (9,9) -- the Polar Arc, M0.16-0.10, and M-0.32-0.19 -- are not strong submillimeter or millimeter sources (red circles in Figure \ref{map}). The latter two of these clouds were studied by \cite[][hereafter RF01]{RF01}, who measure strong emission from the S(3), S(4) and S(5) rotational lines of H$_2$, and derive rotational temperatures of $\sim$600 K.  There are several possible reasons why we might not detect these clouds in highly-excited lines of \am, one of which is that these clouds are not sufficiently dense to give rise to strong \am\, emission lines (to observe \am\, in emission requires densities of $\gtrsim10^3$ \cm). As RF01 find it most likely that these higher H$_2$ lines arise in very dense gas (n $\sim 10^6$ \cm), this is unlikely to be the explanation.  

Another possibility is then that the column density of \am\, in these two clouds is too low for highly-excited lines of \am\, to be detectable. In one of these clouds, M-0.32-0.19, \cite{Huttem93b} also observe \am\, lines from (1,1) to (6,6) and find evidence for a warm gas component with a temperature of $\sim$ 150 K. However, compared to M0.25+0.01, in which we do detect more highly-excited lines of \am,  the column density of \am\, in M-0.32-0.19 is much lower: only 1/5 of the M0.25+0.01 column in the $J$=5 level. If the hot \am\, column in M-0.32-0.19 is also 1/5 that of M0.25+0.01, this is consistent with the upper limits implied by our nondetection.  

We also marginally detect \am\, (8,8) and (9,9) in M0.83-0.10, which favors the idea that the H$_2$ and \am\, may trace the same hot gas, but that the column density of \am\, in many of these clouds is below our detection limit. RF01 find that their hot component makes up $\lesssim$1\% of the column density of warm (T$\sim 150$ K) H$_2$ in these clouds (which is itself only $\sim$25-30\% of the total H$_2$ column), compared to our finding that the hot \am\, component in M-0.02-0.07 makes up $\sim$10\% of the total \am\, column density in this cloud. 
 If the hot gas in M0.83-0.10 has the same properties as that we detect in M-0.02-0.07, we can use the comparison of its (8,8) column density to that measured for M-0.02-0.07 and obtain an estimate of the total \am\, column density in M0.83-0.10. Assuming that 10\% of this gas arises in a T$\sim$ 400 K component, then the column density of hot \am\, in this cloud would be $\sim5\times10^{14}$ cm$^{-2}$,  and compared to the column density of hot gas RF01 detect in this cloud would yield [\am]/[H$_2$] $\sim2\times10^{-6}$. This is consistent with the \am\, abundance \cite{Cecca02} find for the hot \am\, envelope of Sgr B2. It then seems most likely that the clouds we observed which were detected by RF01, while having a relatively high fractional abundance of \am, simply have an \am\, column density which is too low to be detected by our survey.

Finally, there is one cloud in common between our sample and that of RF01(M-0.13-0.08, referred to as M-0.15-0.07 in RF01) for which we do measure a hot gas component of $\sim$400 K, but RF01 do {\it not} detect a high-temperature gas component. RF01 only detect the two lowest H$_2$ rotational lines, S(0) and S(1), in this cloud, from which they measure an intermediate-temperature component of $\sim$ 140 K. The fact that hot \am\, is detected in this cloud, but not hot H$_2$, could either indicate that this cloud has a higher relative \am\, abundance, or that the hot \am\, and hot H$_2$ arise in different environments, and may be heated by separate mechanisms.

\subsection{Density}
\label{densities}

The critical density of the highest-excitation \am\, inversion lines, assuming a kinetic temperature of 300 K and extrapolating the expected collisional rates from the values calculated for transitions up to (6,6) by \cite{Danby88}, is a few $\times10^3$ \cm. Thus, to detect these \am\, transitions in emission requires gas that is at least this dense. The morphology and kinematics of the hot \am\, emission mapped in M-0.02-0.07 and M-0.13-0.08 are also similar to those of the HC$_3$N 3-2 line, observed simultaneously, which has a critical density of n = $10^4$ cm$^{-3}$. However, if the hot \am\, has a substantially higher density than $10^4$ cm$^{-3}$, there are two issues. First, one might expect this dense gas to arise in more compact sources, which would be inconsistent with our finding that $\sim$ 96\% of the hot \am\, emission is resolved out in our VLA observations. Gas in the cloud interior which is denser than $\sim 10^5$ \cm\, should also reach thermal equilibrium with dust in the cloud \citep{Goldsmith01,Juvela11}. As no hot dust is detected from these clouds \citep{Molinari11,Longmore}, this also favors a density lower than $\sim 10^5$ \cm.

The density of the similarly-hot \am\, component in Sgr B2 was determined by a large-velocity-gradient model to be $< 10^4$ cm$^{-3}$ \citep{Cecca02}, which is consistent with the density we infer for the hot gas in our cloud sample.  Though the \am\, in Sgr B2 is hotter (600-700 K, compared to 400-500 K in the clouds we study), the emitting region appears to have a similar size to those we detect  in M-0.02-0.07 and M-0.13-0.08. \cite{Mauers86} mapped the slightly lower-excitation \am\, (7,7) line in Sgr B2, measuring extended emission over a $\sim$ 1.5 arcminute area. For comparison, our maps of (8,8) and (9,9) toward M-0.02-0.07 and M-0.13-0.08 show that the emitting region has a size of $\sim 1.5'-2'$. However, we note that the Sgr B2 cloud is a very actively star forming cloud which is an order of magnitude more massive than other CMZ clouds, and its properties are not generally typical of other Galactic center molecular clouds.

There are also three clouds in our sample for which RF01 derive the physical conditions from pure-rotational lines of H$_2$, finding T $\sim$ 600 K and n $\sim10^6$ \cm. We marginally detect emission from the (8,8) and (9,9) lines in one of these clouds (M0.83-0.10), and suggest that for the others, the column density of hot \am\, is below our detection limits. 
  If the hot \am\, and hot H$_2$ arise in the same gas component, this suggests that the hot \am\, does arise in gas denser than 10$^{5}$ \cm, in which case, it is not clear why this gas is not in thermal equilibrium with the dust in these clouds. However, the RF01-derived temperatures and densities are based on purely collisional excitation of H$_2$; if there is additional radiative excitation, the actual H$_2$ densities and temperatures could be lower, possibly avoiding this discrepancy. 

\begin{figure*}
\includegraphics[scale=0.4]{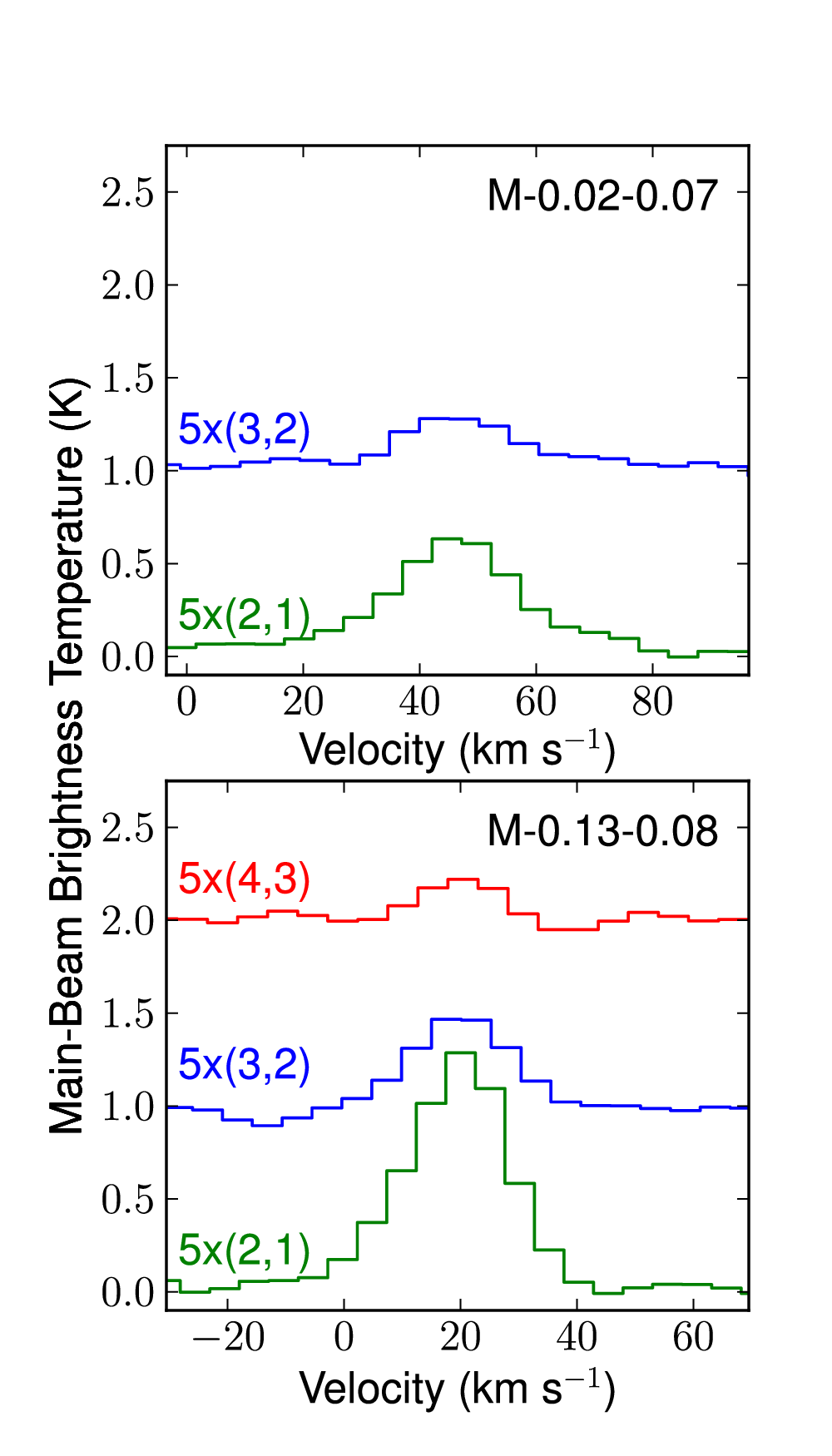}
\caption{ Spectra of non-metastable \am\, lines toward M-0.02-0.07 and M-0.13-0.08.}
\label{nonmeta}
\end{figure*}

 We also detect non-metastable lines of \am\, in both M-0.02-0.07 and M-0.13-0.08, the only clouds in our sample for which we observed these lines (Figure \ref{nonmeta}). The presence of non-metastable \am\, implies that either the emitting gas is quite dense, or that there is a strong far-infrared radiation field \citep{Zuckerman71,Sweitzer79}. However, the density of this gas may be unrelated to the hot \am\, component we observe, as \cite{Huttem93b} showed that the lower-excitation \am\, lines trace a separate, cool gas component of T$\sim$ 25 K. Further, the nonmetastable and metastable emission even from lower-excitation lines can arise from different regions of the cloud. In Sgr B2, maps of both nonmetastable and metastable lines show that the nonmetastable \am\, emission is clearly spatially distinct from the emitting region of the lower-excitation metastable \am\, lines \citep{Huttem93}. 
	
\subsection{Heating}
\label{heating}
	
	We first consider whether the \am\, emission might be nonthermal: the result of formation in a highly-excited state (a phenomenon known as formation pumping). In this case, the highly-excited \am\, which we observe would be the result of the incomplete thermalization of these molecules after their formation. Formation at high temperatures implies an ortho/para ratio of 1, largely consistent with our observations \citep{Tak02}. However, our determination that $\sim$10\% of the \am\, column in M-0.02-0.07 is hot also puts a tight constraint on the \am\, formation rate. Assuming that it takes $\sim$ 5-10 collisions for an \am\, molecule to reach thermal equilibrium with the surrounding gas after it has been formed, and given a collision rate coefficient of $\sim 2 \times 10^{-10}$ cm$^3$ s$^{-1}$ and a gas density of at least $10^3$ \cm, we find that an \am\, molecule will come to thermal equilibrium in $\sim$ 1 year. This means that 10\% of the \am\, in these clouds must have formed within a few years. Thus in a steady state (inferred from the fact that hot ammonia is observed to be widespread throughout the GC), the typical lifetime of an \am\, molecule in the gas phase in CMZ clouds would only be $\sim$ 10 years.  Looking at this another way, if \am\, must be destroyed at the same rate at which it is formed, then every hundredth collision an \am\, molecule experiences is with a molecule (like H$_3^+$) which destructively reacts with the \am, which would indicate a highly unlikely abundance of reactants. Although the \am\, could be destroyed on these timescales via photodissociation if it is located in PDRs \citep{RF04},  the authors find this unlikely, as it would require an unreasonably high inflow rate of \am-rich gas into the PDR regions to maintain the observed \am\, abundance. Another possibility, the continuous formation and destruction of \am\, in dissociative (J-type) shocks, has also been ruled out by \cite{RF04}, who find that J-type shock models do not reproduce the observed fine-structure line intensities in the GC. We thus conclude that the highly-excited \am\, we observe is likely to be due to a thermal and truly hot component of CMZ clouds.

The widespread distribution of highly-excited \am\, in cloud throughout the CMZ favors a global heating source, such as shocks, cosmic rays, or possibly X-rays. Cosmic rays \citep{YZ07, Goto08}, and the dissipation of turbulence (possibly magnetohydrodynamic turbulence) via shocks \citep{MA75,Wilson82,MP97}, have both been suggested to be possible sources of GC heating. Both also can result in gas temperatures systematically higher than dust temperatures, an observed property of clouds in the CMZ \citep{Molinari11}. RF01 also find that their hot (T$\sim$600 K) H$_2$ component is consistent with either models of dense PDRs or shocks, though the exact relationship between this component and the 400 K \am\, presented here is unclear.

It is possible that PDRs do play some role in heating these clouds. In our temperature maps of M-0.02-0.07, we noticed consistently higher temperatures on the southeast edge of this cloud, a location which roughly corresponds with a group of 4 compact \hii regions embedded in the cloud.  However, while the embedded \hii regions may make some local contribution to the heating of this cloud, we note that many of the clouds in which we detect highly-excited \am\, do not have embedded \hii regions or associated star formation (e.g., M0.25+0.01, M0.11-0.08, M0.34+0.06). This suggests that heating due to embedded sources is not the dominant heating mechanism for this hot \am\,. 

 The emission from these highly-excited \am\, lines also appears concentrated toward the cloud centers, and is thereby unlikely to arise in the cloud envelopes (see the discussion in Section \ref{distribution}). This argues against the heating occurring in an irradiated outer layer, though a penetrating heating source, such as cosmic rays, could be plausible. However, recent modeling of a typical GC cloud \citep[M0.25+0.01;][]{Clark13} suggests that cosmic ray heating, even for relatively high cosmic-ray ionization rates of $\zeta\sim$ a few $10^{-14}$ s$^{-1}$, \citep[e.g., those recently inferred by ]{YZ07b,YZ13b,YZ13c}, does not heat even the most tenuous gas (n$\sim10^3$ \cm) above 300 K, which is less than the lower limits on the kinetic temperature we derived from the rotational temperatures of the three clouds studied (M0.25+0.01, M-0.02-0.07, and M-0.13-0.08). Finally, as the emission is also largely resolved out by interferometric observations with the VLA, it must be extended on spatial scales of tens of arcseconds and distributed relatively uniformly throughout the cloud. This could then either occur if the emission originates in structures too narrow to resolve with VLA (such as shock fronts), or if the emission emanates from a generally smooth component, such as a less-dense inter-clump medium. 
 
\begin{figure*}
\hspace{1cm}
\includegraphics[scale=0.9]{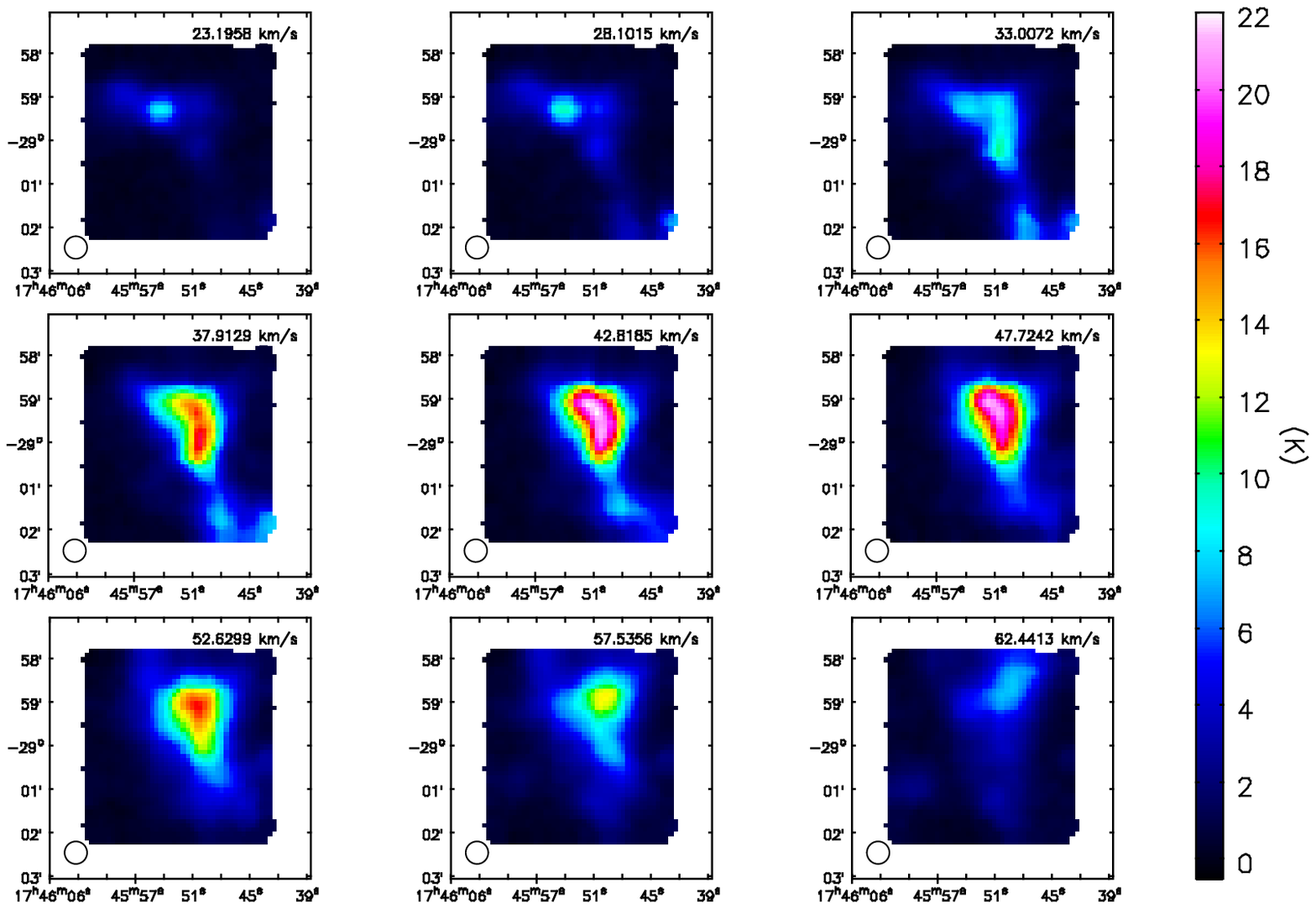}
\vspace{1cm}
\caption{Channel maps of \am\, (3,3) emission in M-0.02-0.07. }
\label{chan_33}
\end{figure*}

\begin{figure*}
\hspace{1cm}
\includegraphics[scale=0.2]{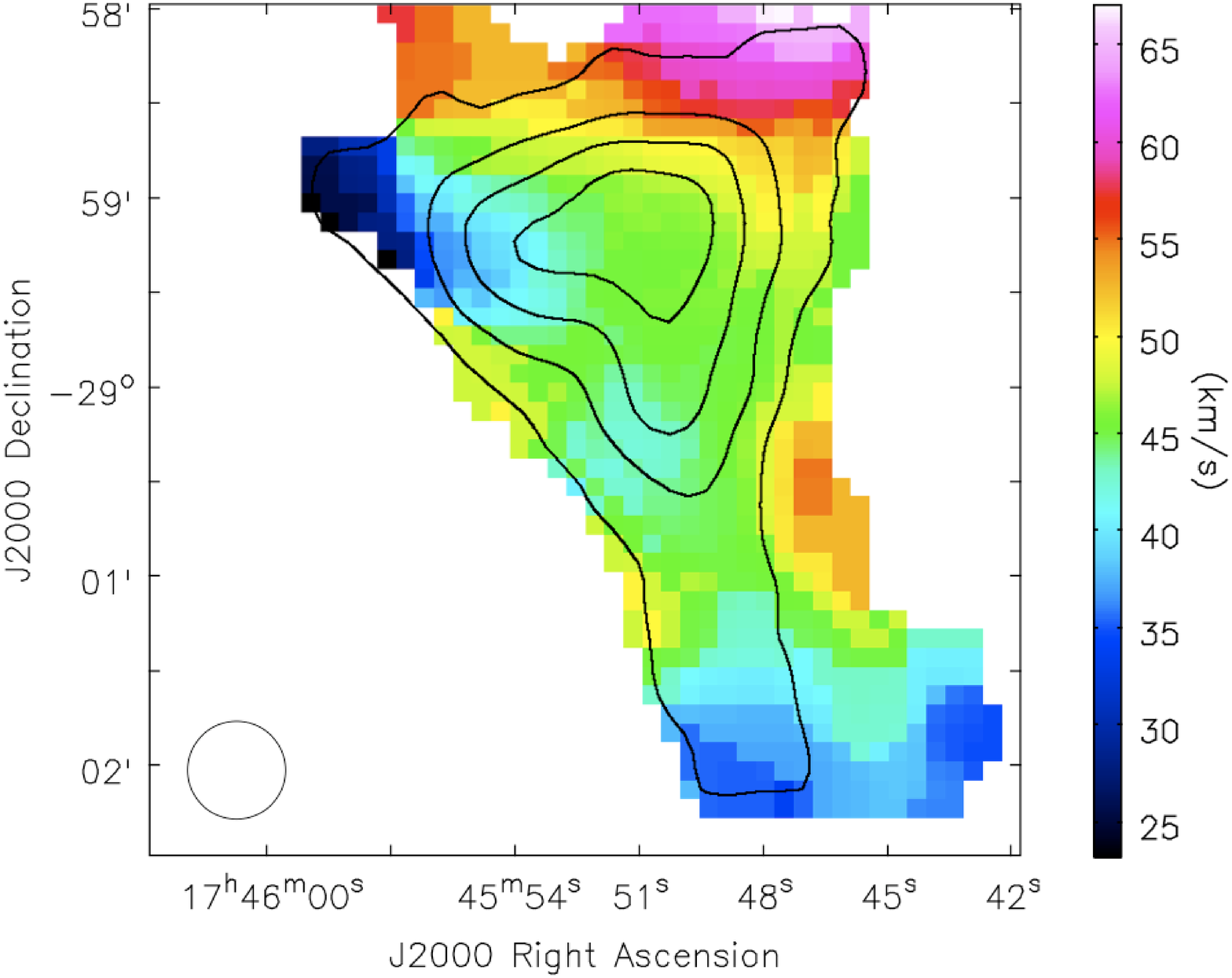}
\hspace{1cm}
\includegraphics[scale=0.2]{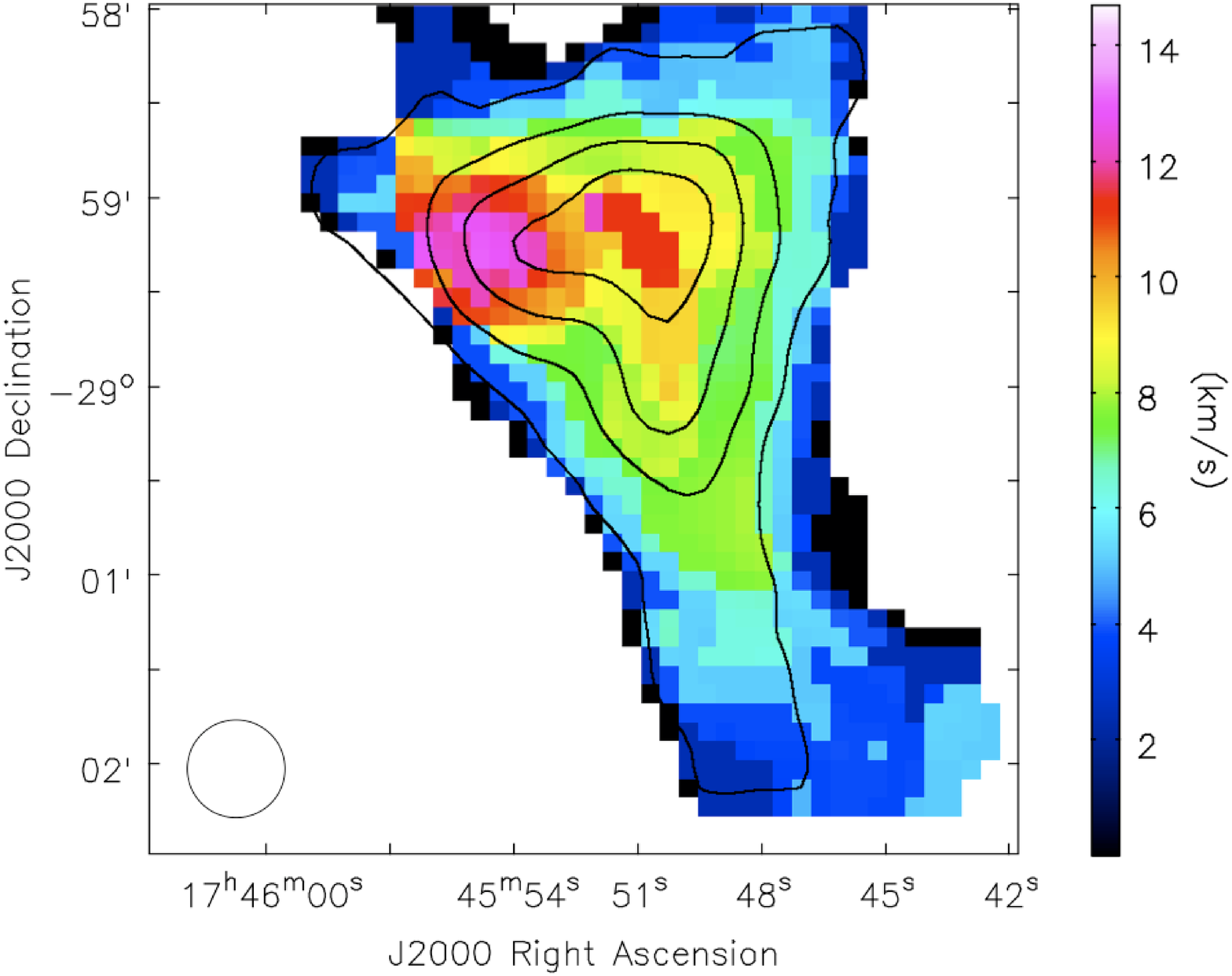}
\vspace{1cm}
\caption{{\bf Left:} Intensity-weighted velocity in M-0.02-0.07, from the the \am\, (3,3) line. {\bf Right:} Intensity-weighted velocity dispersion in M-0.02-0.07, from the the \am\, (3,3) line. Contours of emission from the \am\, (6,6) line are overplotted on both maps in black. }
\label{mom_33}
\end{figure*}

We briefly investigate the role of cloud-cloud collisions on heating in M-0.02-0.07 by comparing the temperature maps with the kinematic structure of the cloud. Figure \ref{chan_33} shows the channel maps of the ammonia (3,3) line, the strongest line in this cloud, and thus the best tracer of the underlying kinematic structure. This cloud exhibits emission over a relatively wide range of velocities, from 20 to 70 \kms. In the righthand side of Figure \ref{mom_33}, we show an intensity-weighted velocity map of the cloud, also using the (3,3) line. Here, one can see two prominent velocity gradients: a north-south gradient from 40 to 65 \kms on the western edge of the cloud, corresponding to the ridge of material swept up by the supernova shell Sgr A east, and an east-west gradient from 25 to 45 \kms, where the eastern extension of the cloud appears to join this swept-up shell. The area where these velocity components blend is host to a large number of collisionally-excited (Class I) CH$_3$OH masers \citep{SJ10,P11}, indicative of strong shocks. We hypothesize that the higher temperatures on the southeast edge of M-0.02-0.07 which are seen in all of the maps may actually not be due to the embedded \hii regions, but instead to the collision of these two components of the cloud. The strong temperature peak on the northern edge of the cloud seen in the (4,4)-(8,8) and (6,6)-(9,9) temperature maps could also be related to this interaction. If large-scale collisions do contribute to variations in temperature in the hot gas component we observe, then other clouds with strong emission from highly-excited lines of \am\, such as M0.25+0.01, M-0.13-0.08, and M0.11-0.08, should also show a correlation between their temperature structure and their kinematics, as well as emission from shock-tracers such as Class I CH$_3$OH masers. 

Finally, we also compare the velocity dispersion in M-0.02-0.07 with its temperature. The lefthand side of Figure \ref{mom_33} shows the velocity dispersion of the (3,3) line. The velocity dispersion is generally higher toward the center of the cloud, and peaks at the eastern edge. It is possible that this region of larger velocity dispersion is related to the slightly enhanced temperatures in this area of the (3,3) to (6,6) rotational temperature map (Figure \ref{temp_all}). However, the other two temperature maps, which are derived from more highly-excited lines of \am, show no correlation of temperature with an increased turbulent line width. 
 
There are several potential methods for further distinguishing between the different mechanisms for heating the hot gas component we detect in CMZ clouds. With higher-resolution interferometric observations of highly-excited \am\, (for example, in absorption against \hii regions) one could test whether the distribution of highly excited \am\, varies on the small scales which would be consistent with arising in narrow shocks. One could also examine a larger sample of clouds for a correlation between higher temperatures or a larger fraction of hot gas and tracers of enhanced shock activity in these clouds, for example, collisionally-excited CH$_3$OH masers which have been found to be extremely abundant throughout the CMZ  \citep{YZ13}. \cite{Ao13} also suggest several means to distinguish between cosmic rays and turbulent shocks as the heating source for CMZ clouds. They suggest both searching for an enhanced ionization fraction which would be indicative of a high cosmic ray ionization rate, and looking for hot, low-velocity-dispersion structures within individual clouds, which would indicate that the dissipation of turbulent energy is not the primary source of cloud heating. 

\section{Conclusions}

Based on our detections of \am\, lines with energy levels up to 2200 K above the ground state, this paper presents the following findings:

1. We have detected a hot gas component associated with a large number of Galactic center molecular clouds. Of the 17 positions we surveyed for \am\, emission, we detect \am\, (9,9) toward 13 positions, and marginally detect it toward two positions. For those clouds in which we observed only (8,8) and (9,9), we estimate T$>200-300$ K. 

\vspace{0.2cm}
2. For three clouds, M0.25+0.01, M-0.02-0.07, and M-0.13-0.08,  we have detected emission from lines of \am\, up to (15,15) and measure temperatures of T $\sim$ 350-450 K. A similar, though hotter, gas component was previously known in Sgr B2 \citep[T=600-700 K;][]{Cecca02,Wilson06}, but the observations presented here indicate that hot gas is widespread in both lower mass and more quiescent GC clouds throughout the central 300 parsecs.  

\vspace{0.2cm}
3. For M-0.02-0.07, we have also mapped the \am\, (9,9) emission with the VLA, and find that, compared to our GBT observations,  $\sim$96\% of the flux from this line is resolved out. This indicates that the hot \am\, emission is uniform on $\sim$ 10 arcsecond scales, either because it arises in a smooth, extended component, or in a network of unresolved structures, such as shocks.

\vspace{0.2cm}
4. We have also measured the \am\, (1,1) through (6,6) lines in M-0.02-0.07, allowing us to determine that the hot (T $\sim$ 400 K ) \am\,  in this cloud contributes 7-12\% of the total \am\, column density. This makes it unlikely that the highly-excited \am\, lines are a result of formation pumping-- a nonthermal population of the \am\, levels due to formation in a highly-excited state. 

\section{Acknowledgements}
We wish to thank the Green Bank Telescope support staff, and especially to thank Ron Maddalena for extremely helpful discussions with regards to the calibration of the data. We also wish to thank T. Oka and D. Meier for their useful comments on the manuscript, and we are grateful to the anonymous referee for the careful and insightful suggestions which improved the final version.
This material is based upon work supported by a grant from the NRAO Student Observing Support program. 

\bibliographystyle{hapj}
\bibliography{ammonia.bib}

\appendix
\section{Properties of clouds for which we detect \am\, (9,9)}
\label{individual}

\subsubsection{G1.6-0.025}
G1.6-0.025 is a cloud at the extreme high-longitude end of the CMZ. This cloud exhibits remarkably strong lines of CH$_3$OH, indicative of intense shocks \citep{Menten09}.  We detect relatively strong emission from the (9,9) line in this cloud, but do not detect the (8,8) line, which is surprising given the apparent strength of the (9,9) line. 
Because of the discrepant strength of the (9,9) line, G1.6-0.025 is one of two clouds which we suggest in Section \ref{o_p} may have an anomalous ortho to para \am\, ratio (O/P $>$ 1.5). This ratio, which is larger than that measured toward the clouds in Table \ref{temps}, indicates that the \am\, in this cloud may have formed at cooler temperatures (or if it formed via grain surface reactions, that it reflects cooler grain temperatures in this cloud). 

\subsubsection{M0.83-0.10}
M0.83-0.10 lies to the southeast of Sgr B2. The temperature of this cloud has previously been measured by \cite{RF01} using pure-rotational lines of H$_2$. They find two temperature components: a warm component, with T $\sim$ 180 K, and a hot component, with T$\sim$ 600 K. The warm component is estimated to make up 25\% of the total H$_2$ column density in this cloud. We marginally detect the \am\, (8,8) and (9,9) lines toward this source.

\subsubsection{Sgr B2}
We detect \am\, (8,8) and (9,9) in absorption against the continuum from the hot cores Sgr B2-M and Sgr B2-N. Higher lines of \am\, have previously been studied in Sgr B2 \citep{Huttem93,Flower95, Cecca02,Wilson06}, yielding rotational temperatures in excess of 700 K. The Sgr B2 cloud has an extremely high \am\, column density, leading to suggestions that the \am\, abundance is enhanced, perhaps by shocks \citep{Cecca02}. We measure N/T$_{\rm ex}$ $\sim$ 10$^{13}$ cm$^{-2}$, comparable to that measured by \cite{Huttem95} and \cite{Wilson06}. \cite{Wilson06} estimate that the \am\, excitation temperature varies from 6 K for \am\, (1,1) to 3 K for \am\, (18,18), so the total \am\, column densities in these lines are likely $\lesssim$ a few $\times 10^{13}$.

In our spectra of \am\, (9,9) toward these sources we also detect the $J_K$= 13$_2$-13$_1$ line of CH$_3$OH, with a rest frequency of 27.47253 GHz. The rest frequency of this line is very similar to that of the (9,9) line at 27.477943, and it appears in emission at a velocity offset of $\sim$120 \kms\, in the (9,9) spectra of both Sgr B2-N and Sgr B2-M, in Figure \ref{detect}.  That this is not an \am\, emission feature is apparent both from this velocity offset, which is inconsistent with the radial velocities of many species from these sources, and from the fact that no corresponding emission feature is observed in the (8,8) spectra. Additional $J_K$=$J_2$-$J_1$ transitions of CH$_3$OH have also been previously detected toward both sources \citep{Pei00}.

\subsubsection{M0.34+0.06}
M0.34+0.06 is one of the clouds in the `Dust Ridge' that stretches from M0.25+0.01 to Sgr B2 \citep{Lis94}. More recently, these clouds have been interpreted as representing gas on x2 orbits in the Galactic center potential \citep{Molinari11}. Although several clouds in the dust ridge have associated 6.7 GHz (Class I or radiatively-excited) CH$_3$OH masers \citep{Caswell10}, which imply the presence of embedded star formation, this cloud shows no indication of ongoing star formation. Emission from highly-excited lines of  \am\, in this cloud is relatively faint, with linewidths $\sim$ 20 \kms. 

\subsubsection{M0.25+0.01}
M0.25+0.01 is another apparently quiescent cloud in the Dust Ridge: the only evidence for ongoing star formation in this cloud is a single, weak water maser \citep{Lis94}. Linewidths in this cloud vary from 15 to 20 \kms. The cloud is quite massive (M $\sim10^5$M$_{\odot}$) and has a very high total column density (N$_{H_2}\sim4\times10^{23}$ cm$^{-2}$). As it appears prominently in absorption in infrared images, it likely lies toward the near edge of the CMZ \citep{Longmore}. Although this cloud has previously been measured to be relatively cold (\cite{Huttem93b} measure \am\, temperatures of 17 and 80 K in this cloud, and \cite{Ao13} find a temperature of $\sim$ 70 K using lines of H$_2$CO ), we find that this cloud also has a hot gas component of $\sim$400 K. 

\subsubsection{G0.18-0.04}
G0.18-0.04 is an \hii region (the Sickle) and an intense photodissociation region (PDR) on the face of a molecular cloud adjacent to the massive Quintuplet star cluster \citep{YZ87,Nagata90}. Although the cloud is detected in relatively highly-excited 3--2 and 5--4 transitions of CS \citep{Serabyn91}, we detect only faint \am\, emission from the (9,9) line and marginally detect the (8,8) line toward this cloud. Compared to M-0.02-0.07, for which the ratio of CS 5--4 (measured by \cite{Serabyn92}) to \am\, (9,9) is $\sim$5, the ratio in this cloud is 20. This suggests that either this cloud, as a strong PDR, has a smaller fraction of hot molecular gas than other clouds we survey, or that the hot \am\, is located in a less-shielded region of this cloud than the CS where it is preferentially destroyed, as both molecules have similar photodissociation rates \citep{Martin12,Roberge81}. 

\subsubsection{G0.07+0.04}
The G0.07+0.04 cloud is also associated with an intense PDR region, and its ionized surface forms part of the thermal arched filaments complex \citep{Serabyn87,Lang01,Lang02}, which is ionized by the massive Arches star cluster \citep{Nagata95, Cotera96}. It is prominent in maps of the CII cooling line \citep{Pog91}. Similar to G0.18-0.04, the \am\, (9,9) emission we detect from this cloud is very faint, despite both sources being clouds with dense molecular gas immersed in a PDR, an environment in which one might expect to find hot gas. Again, this suggests either that hot \am\, is not associated with PDRs, and so PDRs are not a strong source of this emission, or that hot \am\, does exist in PDRs, but is depleted in the most intense such environments.  %From their observations of high ammonia abundances, RF04 argues that \am\, in Galactic center clouds is not associated with PDRs, as it should be entirely destroyed on timescales of < 10 years in the typical Galactic center radiation field (G$_0\sim10^3$). %If the hot \am\, emission is associated with PDRs, then it should originate in a low-opacity, less shielded region of the cloud, and not in dense interior clumps. 

\subsubsection{M0.11-0.08}
M0.11-0.08 is one of several prominent molecular cloud cores just to the east of Sgr A and is a relatively bright continuum source at millimeter and submillimeter wavelengths \citep{PP00,Bally10} as well as as in several molecular tracers  \citep{Gusten81, AB85}.  It is also part of the larger M0.11-0.11 complex which is notable for having a very high column density of SiO \citep{Tsuboi,Handa}.  We find that M0.11-0.08 has one of the highest column densities of \am\, (9,9) in our sample, next to M-0.13-0.08 and M-0.02-0.07 (and Sgr B2, which we observe only in absorption).  \cite{Ao13} also found that this was the second hottest cloud in the central 100 parsecs (T$\sim$125 K) as measured with millimeter transitions of formaldehyde. There are no known signs of star formation associated with this cloud.

\subsubsection{M-0.02-0.07}
M-0.02-0.07 is one of two giant molecular clouds in the Sgr A complex, lying on the eastern edge of the Sgr A East supernova remnant, which is apparently compressing the edge of this cloud \citep{Serabyn92}. There are also four compact \hii regions that appear to be partially embedded within this cloud \citep{Goss85,YZ10,Mills11} . The \am\, (9,9) line emission from this cloud is the strongest of all the positions we observed, from which we infer that this cloud has the highest hot \am\, column density of all clouds in our sample, next to Sgr B2. \cite{Ao13} found that this was the hottest cloud in the central 100 parsecs (T $\sim$ 190 K) as measured with millimeter transitions of formaldehyde. We discuss this cloud more in Section \ref{distribution}.

\subsubsection{The Circumnuclear disk / Southern streamer }
The Circumnuclear disk (CND) is a ring of gas and dust surrounding the central supermassive black hole at a radius of 1.5 pc \citep{Becklin82,Genzel85}. Lower-excitation lines of \am\, up to (6,6) have been detected in the CND \citep{McGary01,HH02}, and we marginally detect emission from \am\, (8,8) and (9,9) at $\sim$60 \kms. This velocity is consistent with the previous \am\, spectra of the CND toward this position, which is offset from Sgr A* by (R.A, Dec.) =  (-6$''$,-39$''$). However, because the broad width of these lines are comparable to the width of fluctuations in the surrounding baseline of the spectra, their intrinsic shape is uncertain, and we are unable to robustly fit for the properties of these lines.

We also detect both the (8,8) and (9,9) lines in absorption, likely against the `minispiral' of Sgr A \citep{Ekers83,Zhao09}, at a velocity of 35 \kms.  This gas is likely part of the `Southern streamer', identified by  \cite{Coil99} and \cite{Coil00} as an extension of M-0.13-0.08 that may interact with the CND.

\subsubsection{M-0.13-0.08}
M-0.13-0.08 is the second giant molecular cloud in the Sgr A complex, lying to the west of Sgr A West and the CND. Like M-0.02-0.07, this cloud shows some evidence for recent star formation, as it has a single embedded \hii region \citep{Ho85}. The cloud also appears to be interacting with a supernova remnant on its periphery \citep{Coil00}. After M-0.02-0.07, this cloud has the strongest high-excitation \am\, emission line strengths in our sample. 

\subsubsection{Sgr C} 
As the distribution of targets in our sample follows the distribution of dense gas in the CMZ, which is predominantly located to the east of Sgr A* at positive latitudes \citep{Morris96}, we observed relatively few positions at negative latitudes. Of the four positions we observed that lie to the east of Sgr A*, Sgr C is the only one aside from M-0.13-0.08 in which we detect \am\, (9,9). The line emission is, however, quite weak.

\clearpage

\section{Figures and Tables}

\begin{table}[ht]
\caption{Observed Sources} 
\centering
\begin{tabular}{cccrr}
\\[0.5ex]
\hline\hline
& & & & \\
  &  & & Galactic & Galactic \\
{\bf Source} & {\bf RA (J2000)} &{\bf Dec (J2000)} & Longtitude & Latitude\\
\hline
 {\bf G1.6-0.025            } & 17$^{\mathrm{h}}$49$^{\mathrm{m}}$19.9$^{\mathrm{s}}$ & -27\degr 34\arcmin 11.0\arcsec &   1\degr.376 & -0\degr.122  \\
 {\bf M0.83-0.10            } & 17$^{\mathrm{h}}$47$^{\mathrm{m}}$57.9$^{\mathrm{s}}$ & -28\degr 17\arcmin 00.0\arcsec &   0\degr.383 & -0\degr.368 \\
 {\bf Sgr B2 (N)            } & 17$^{\mathrm{h}}$47$^{\mathrm{m}}$20.1$^{\mathrm{s}}$ & -28\degr 22\arcmin 21.0\arcsec &   0\degr.321 & -0\degr.244 \\
 {\bf Sgr B2 (M)            } & 17$^{\mathrm{h}}$47$^{\mathrm{m}}$20.3$^{\mathrm{s}}$ & -28\degr 23\arcmin 06.0\arcsec &   0\degr .321 & -0\degr.245 \\
 {\bf Dust ridge            } & 17$^{\mathrm{h}}$46$^{\mathrm{m}}$11.5$^{\mathrm{s}}$ & -28\degr 37\arcmin 10.0\arcsec &   0\degr.207 & -0\degr.020 \\
 {\bf M0.25+0.01            } & 17$^{\mathrm{h}}$46$^{\mathrm{m}}$10.3$^{\mathrm{s}}$ & -28\degr 43\arcmin 37.0\arcsec &   0\degr .205 & -0\degr.016 \\
 {\bf G0.18-0.04            } & 17$^{\mathrm{h}}$46$^{\mathrm{m}}$14.0$^{\mathrm{s}}$ & -28\degr 46\arcmin 49.0\arcsec &   0\degr.211 & -0\degr.028 \\
 {\bf M0.16-0.10            } & 17$^{\mathrm{h}}$46$^{\mathrm{m}}$26.8$^{\mathrm{s}}$ & -28\degr 51\arcmin 05.0\arcsec &   0\degr.233 & -0\degr.070 \\
 {\bf M0.11-0.08            } & 17$^{\mathrm{h}}$46$^{\mathrm{m}}$13.3$^{\mathrm{s}}$ & -28\degr 53\arcmin 29.0\arcsec &   0\degr.210 & -0\degr.026 \\
 {\bf Polar Arc             } & 17$^{\mathrm{h}}$45$^{\mathrm{m}}$06.6$^{\mathrm{s}}$ & -28\degr 46\arcmin 27.0\arcsec &   0\degr.099 & +0\degr.191 \\
 {\bf G0.07+0.04            } & 17$^{\mathrm{h}}$45$^{\mathrm{m}}$37.5$^{\mathrm{s}}$ & -28\degr 52\arcmin 40.0\arcsec &   0\degr.151 & +0\degr.091 \\
 {\bf M-0.02-0.07           } & 17$^{\mathrm{h}}$45$^{\mathrm{m}}$52.4$^{\mathrm{s}}$ & -28\degr 59\arcmin 02.0\arcsec &   0\degr.175 & +0\degr.042 \\
 {\bf Southern Streamer/CND } & 17$^{\mathrm{h}}$45$^{\mathrm{m}}$39.5$^{\mathrm{s}}$ & -29\degr 01\arcmin 07.0\arcsec & 359\degr.300 & -0\degr.437  \\
 {\bf M-0.13-0.08-b         } & 17$^{\mathrm{h}}$45$^{\mathrm{m}}$37.9$^{\mathrm{s}}$ & -29\degr 03\arcmin 52.0\arcsec & 359\degr.298 & -0\degr.432 \\
 {\bf M-0.13-0.08           } & 17$^{\mathrm{h}}$45$^{\mathrm{m}}$37.4$^{\mathrm{s}}$ & -29\degr 05\arcmin 37.0\arcsec & 359\degr.297 & -0\degr.431 \\
 {\bf M-0.32-0.19           } & 17$^{\mathrm{h}}$45$^{\mathrm{m}}$37.7$^{\mathrm{s}}$ & -29\degr 18\arcmin 28.0\arcsec & 359\degr.297 & -0\degr.431 \\
 {\bf Sgr C                 } & 17$^{\mathrm{h}}$44$^{\mathrm{m}}$40.7$^{\mathrm{s}}$ & -29\degr 27\arcmin 59.0\arcsec & 359\degr.203 & -0\degr.247 \\
& & & &  \\
\hline

\end{tabular}
\label{Sources}
\end{table}

\begin{table}[ht]
\caption{Observed Transitions of \am} 
\centering
\begin{tabular}{cccc}
\\[0.5ex]
\hline\hline
& & &  \\
Transition & Frequency & Upper State Energy\footnotemark[1] & GBT beam \\ [0.5ex]
\hline
{\bf ($J,K$)} & {\bf (MHz)} & {\bf (K)} & {\bf FWHM}\\
(4,3)      & 22688.3120 & 237.8 & 33.3$''$ \\
(3,2)      & 22834.1851 & 150.2 & 33.0$''$ \\
(2,1)      & 23098.8190 & 80.4 & 32.7$''$ \\
(1,1)      & 23694.4955 & 23.3 & 31.8$''$ \\
(2,2)      & 23722.6333 & 64.4 & 31.8$''$ \\
(3,3)      & 23870.1292 & 123.5 & 31.6$''$ \\
(4,4)      & 24139.4163 & 200.5 & 31.3$''$ \\
(6,6)      & 25056.0250 & 408.1 & 30.1$''$ \\
(8,8)      & 26518.9810 & 686.8 & 28.4$''$ \\
(9,9)      & 27477.9430 & 852.8 & 27.5$''$ \\
(10,10) & 28604.7370 & 1036.4 & 26.4$''$ \\
(11,11) & 29914.4860 & 1237.6 & 25.2$''$ \\
(12,12) & 31424.9430 & 1456.4 & 24.0$''$ \\
(13,13) & 33156.8490 & 1692.7 & 22.8$''$ \\
(15,15) & 37385.1280 & 2217.2 & 20.2$''$ \\
& &  \\
\hline
\footnotetext[1]{From the JPL Submillimeter, Millimeter, and Microwave Spectral Line Catalog \citep{JPL}}
\end{tabular}
\label{ammonia}
\end{table}

\clearpage

\begin{table}[ht]
\caption{\am\, Line Parameters}
\centering
\begin{tabular}{ccccccc}
\\[0.5ex]
\hline\hline
& & & & & &  \\
Source & Transition & T$_{\rm mb}$ & v$_{center}$ & $\Delta$v  & Velocity-integrated T$_{\rm mb}$ & Column Density \\ 
              &   & ( K )  & ( km s$^{-1}$ ) &  ( km s$^{-1}$ )&  ( K km s$^{-1}$ )& ( cm$^{-1}$ )  \\ [0.5ex]
\hline
 {\bf G1.6-0.025      } & ( 8, 8) & $<$ 0.08 & & & & \\
 & ( 9, 9) &  0.211$\pm$0.068 & 158.6$\pm$0.8 &  25.3$\pm$1.8 &    5.91$\pm$ 0.41 &  3.7$\pm$0.26 $\times 10^{13}$ \\
\hline
 {\bf M0.83-0.10      } & ( 8, 8) &  0.081$\pm$0.018 &  32.8$\pm$2.2 &  50.3$\pm$5.2 &    4.15$\pm$0.22 &  2.7$\pm$0.15 $\times 10^{13}$ \\
 & ( 9, 9) &  0.084$\pm$0.032 &  &  &   &  \\
 \hline
 {\bf Sgr-B2-M        } & ( 8, 8) & -4.797$\pm$0.153 &  61.9$\pm$0.2 &  19.9$\pm$0.5 & -104.68$\pm$ 5.26 &   \footnotemark[1] 2.5$\pm$0.07 $\times 10^{13}$ \\
 & ( 9, 9) & -5.580$\pm$0.228 &  62.8$\pm$0.2 &  17.5$\pm$0.4 & -103.83$\pm$ 5.26 &   \footnotemark[1] 2.7$\pm$0.06 $\times 10^{13}$ \\
\hline
 {\bf Sgr-B2-N        } & ( 8, 8) & -3.034$\pm$0.612 &  71.1$\pm$0.9 &  27.7$\pm$2.2 &  -82.21$\pm$ 4.84 &   \footnotemark[1] 2.9$\pm$0.23 $\times 10^{13}$ \\
 & ( 9, 9) & -4.391$\pm$0.753 &  71.1$\pm$1.1 &  29.1$\pm$2.6 & -129.62$\pm$ 7.09 &   \footnotemark[1] 4.1$\pm$0.37 $\times 10^{13}$ \\
\hline
 {\bf M0.34+0.06      } & ( 8, 8) & $<$ 0.07 & & & & \\
 & ( 9, 9) &  0.144$\pm$0.051 &   0.8$\pm$1.6 &  20.2$\pm$3.8 &    3.42$\pm$ 0.26 &  2.1$\pm$0.16 $\times 10^{13}$ \\
\hline
 {\bf M0.25+0.01      } & ( 8, 8) &  0.370$\pm$0.016 &  34.2$\pm$0.2 &  19.0$\pm$0.6 &    7.37$\pm$ 0.37 &  4.8$\pm$0.25 $\times 10^{13}$ \\
 & ( 9, 9) &  0.431$\pm$0.034 &  34.6$\pm$0.4 &  21.2$\pm$1.0 &    9.79$\pm$ 0.50 &  6.1$\pm$0.32 $\times 10^{13}$ \\
 & (10,10) &  0.143$\pm$0.028 &  37.9$\pm$0.4 &  16.3$\pm$1.0 &    2.45$\pm$ 0.17 &  1.5$\pm$0.10 $\times 10^{13}$ \\
 & (11,11) &  0.100$\pm$0.023 &  37.9$\pm$0.4 &  15.2$\pm$1.0 &    1.60$\pm$ 0.12 &  9.0$\pm$0.67 $\times 10^{12}$ \\
 & (12,12) &  0.126$\pm$0.022 &  35.5$\pm$0.4 &  19.7$\pm$0.9 &    2.69$\pm$ 0.16 &  1.4$\pm$0.08 $\times 10^{13}$ \\
 & (13,13) &  0.048$\pm$0.013 &  35.8$\pm$0.6 &  13.2$\pm$1.3 &    0.70$\pm$ 0.06 &  3.5$\pm$0.28 $\times 10^{12}$ \\
 & (15,15) & $<$ 0.04 & & & & \\
\hline
 {\bf G0.18-0.04      } & ( 8, 8) & 0.023$\pm$0.008 & & & & \\
 & ( 9, 9) &  0.056$\pm$0.008 &  23.6$\pm$0.5 &  14.0$\pm$1.2 &    0.82$\pm$ 0.05 &  5.1$\pm$0.32 $\times 10^{12}$ \\
\hline
 {\bf M0.16-0.10      } & ( 8, 8) & $<$ 0.08 & & & & \\
 & ( 9, 9) & $<$ 0.10 & & & & \\
\hline
 {\bf M0.11-0.08      } & ( 8, 8) &  0.480$\pm$0.048 &  50.9$\pm$0.4 &  20.8$\pm$0.9 &   10.82$\pm$ 0.58 &  7.1$\pm$0.38 $\times 10^{13}$ \\
 & ( 9, 9) &  0.733$\pm$0.053 &  51.4$\pm$0.3 &  20.5$\pm$0.8 &   16.50$\pm$ 0.85 &  1.0$\pm$0.05 $\times 10^{14}$ \\
\hline
 {\bf Polar-Arc       } & ( 8, 8) & $<$ 0.04 & & & & \\
 & ( 9, 9) & $<$ 0.05 & & & & \\
\hline
 {\bf G0.07+0.04      } & ( 8, 8) &  0.053$\pm$0.010 & -32.9$\pm$0.7 &  19.8$\pm$1.7 &    1.12$\pm$ 0.07 &  7.4$\pm$0.45 $\times 10^{12}$ \\
 & ( 9, 9) &  0.051$\pm$0.010 & -30.3$\pm$0.6 &  22.6$\pm$1.3 &    1.23$\pm$ 0.07 &  7.7$\pm$0.47 $\times 10^{12}$ \\
\hline
 {\bf M-0.02-0.07     } & ( 2, 1) &  0.120$\pm$0.011 &  46.0$\pm$0.6 &  28.4$\pm$1.4 &    3.85$\pm$ 0.20 &  1.6$\pm$0.08 $\times 10^{14}$ \\
 & ( 3, 2) & $<$ 0.05 & & & & \\
  & ( 3, 3) & 19.412$\pm$0.049 &  45.6$\pm$0.4 &  22.7$\pm$1.0 &  533.14$\pm$63.98 &  4.6$\pm$0.55 $\times 10^{15}$ \\
 & ( 4, 4) &  3.312$\pm$0.072 &  45.9$\pm$0.3 &  22.7$\pm$1.1 &   93.19$\pm$11.18 &  7.5$\pm$0.90 $\times 10^{14}$ \\
 & ( 6, 6) &  2.987$\pm$0.125 &  46.2$\pm$0.3 &  22.7$\pm$1.1 &   73.57$\pm$ 8.83 &  5.3$\pm$0.64 $\times 10^{14}$ \\
 & ( 8, 8) &  0.663$\pm$0.021 &  46.3$\pm$0.2 &  25.4$\pm$0.4 &   18.59$\pm$ 0.93 &  1.2$\pm$0.06 $\times 10^{14}$ \\
 & ( 9, 9) &  0.843$\pm$0.029 &  46.6$\pm$0.2 &  24.8$\pm$0.4 &   22.66$\pm$ 1.14 &  1.4$\pm$0.07 $\times 10^{14}$ \\
 & (10,10) &  0.314$\pm$0.033 &  46.8$\pm$0.3 &  21.2$\pm$0.6 &    6.85$\pm$ 0.37 &  4.1$\pm$0.22 $\times 10^{13}$ \\
 & (11,11) &  0.242$\pm$0.021 &  47.5$\pm$0.2 &  21.3$\pm$0.5 &    5.54$\pm$ 0.29 &  3.1$\pm$0.16 $\times 10^{13}$ \\
 & (12,12) &  0.286$\pm$0.015 &  45.5$\pm$0.2 &  26.2$\pm$0.5 &    8.12$\pm$ 0.41 &  4.3$\pm$0.22 $\times 10^{13}$ \\
 & (13,13) &  0.116$\pm$0.013 &  46.9$\pm$0.3 &  21.2$\pm$0.8 &    2.53$\pm$ 0.13 &  1.3$\pm$0.07 $\times 10^{13}$ \\
 & (15,15) &  0.098$\pm$0.017 &  44.2$\pm$0.5 &  21.7$\pm$1.3 &    2.21$\pm$ 0.12 &  9.8$\pm$0.54 $\times 10^{12}$ \\
\hline
 {\bf Streamer        } & ( 8, 8) & -0.136$\pm$0.029 &  34.2$\pm$0.5 &  19.6$\pm$1.2 &   -3.09$\pm$ 0.20 &   \footnotemark[1] 2.8$\pm$0.17 $\times 10^{12}$ \\
 & ( 9, 9) & -0.159$\pm$0.016 &  35.0$\pm$0.3 &  15.0$\pm$0.6 &   -2.52$\pm$ 0.14 &   \footnotemark[1] 3.6$\pm$0.15 $\times 10^{12}$ \\
\hline
 {\bf CND             } & ( 8, 8) & 0.049$\pm$0.02 & & & & \\
 & ( 9, 9) &  0.068$\pm$0.026 & -62.0$\pm$6.3 &  52$\pm$14.8 &    3.78$\pm$ 0.21 &  2.4$\pm$0.13 $\times 10^{13}$ \\
\hline
 {\bf M-0.13-0.08-b   } & ( 8, 8) &  0.290$\pm$0.016 &   8.7$\pm$0.3 &  20.2$\pm$0.6 &    6.44$\pm$ 0.33 &  4.2$\pm$0.22 $\times 10^{13}$ \\
 & ( 9, 9) &  0.343$\pm$0.017 &   8.0$\pm$0.3 &  23.0$\pm$0.7 &    8.74$\pm$ 0.44 &  5.5$\pm$0.28 $\times 10^{13}$ \\
\hline
 {\bf M-0.13-0.08     } & ( 2, 1) &  0.250$\pm$0.013 &  19.6$\pm$0.3 &  20.9$\pm$0.6 &    5.64$\pm$ 0.29 &  2.3$\pm$0.12 $\times 10^{14}$ \\
 & ( 3, 2) &  0.099$\pm$0.019 &  20.1$\pm$0.6 &  18.7$\pm$1.4 &    1.92$\pm$ 0.13 &  3.9$\pm$0.27 $\times 10^{13}$ \\
 & ( 4, 3) &  0.047$\pm$0.012 &  19.8$\pm$0.6 &  14.2$\pm$1.5 &    0.64$\pm$ 0.07 &  9.7$\pm$1.01 $\times 10^{12}$ \\
 & (10,10) &  0.211$\pm$0.025 &  19.8$\pm$0.3 &  15.1$\pm$0.7 &    3.32$\pm$ 0.19 &  2.0$\pm$0.11 $\times 10^{13}$ \\
 & (11,11) &  0.126$\pm$0.009 &  18.7$\pm$0.3 &  18.4$\pm$0.8 &    2.43$\pm$ 0.12 &  1.4$\pm$0.07 $\times 10^{13}$ \\
 & (12,12) &  0.197$\pm$0.014 &  19.4$\pm$0.2 &  18.4$\pm$0.5 &    3.91$\pm$ 0.20 &  2.1$\pm$0.11 $\times 10^{13}$ \\
 & (13,13) &  0.050$\pm$0.016 &  20.1$\pm$0.9 &  15.8$\pm$2.1 &    0.91$\pm$ 0.07 &  4.6$\pm$0.35 $\times 10^{12}$ \\
 & (15,15) &  0.055$\pm$0.013 &  20.9$\pm$0.5 &  15.3$\pm$1.1 &    0.90$\pm$ 0.06 &  4.0$\pm$0.27 $\times 10^{12}$ \\
\hline
 {\bf M-0.32-0.19     } & ( 8, 8) & $<$ 0.07 & & & & \\
 & ( 9, 9) & $<$ 0.09 & & & & \\
\hline
 {\bf Sgr-C           } & ( 8, 8) &  0.064$\pm$0.019 & -53.0$\pm$0.8 &  12.5$\pm$1.9 &    0.77$\pm$ 0.09 &  5.1$\pm$0.56 $\times 10^{12}$ \\
 & ( 9, 9) &  0.089$\pm$0.017 & -52.9$\pm$0.4 &  12.4$\pm$0.9 &    1.18$\pm$ 0.09 &  7.4$\pm$0.56 $\times 10^{12}$ \\
\hline
\footnotetext[1]{Reported quantity is N/T$_{\rm ex}$, in units of cm$^{-2}$ / K.}
\end{tabular}
\label{results}
\end{table}
\clearpage

\begin{table}[ht]
\caption{\am\, Fits to Hyperfine Line Structure for M-0.02-0.07}
\centering
\begin{tabular}{ccccccc}
\\[0.5ex]
\hline\hline
& & & & & &  \\
Transition & Peak T$_{\rm ex}\times\tau$ & v$_{center}$ & $\Delta$v & Velocity-integrated T$_{\rm ex}\times\tau$ & \footnotemark[1]Column Density & Opacity ($\tau$) \\ 
                   & ( K )                                    & ( km s$^{-1}$ ) &  ( km s$^{-1}$ ) &  ( K km s$^{-1}$ )      &( cm$^{-1}$ )  & \\ [0.5ex]
 \hline
 ( 1, 1) &   8.0$^{+1.3}_{-4.0}$ & 47.0 $\pm$ 0.1  & \footnotemark[2]22.7$^{+9.3}_{-1.7}$ & 430$^{+70}_{-220}$ &  5.7$^{+0.9}_{-2.9}\times10^{15}$ & 1.04$^{+0.26}_{-0.94}$\\
 ( 2, 2) & 10.0$^{+8.5}_{-5.2}$ & 46.2 $\pm$ 0.1  & \footnotemark[2]22.7$^{+9.3}_{-5.7}$ & 510$^{+430}_{-260}$ &  5.1$^{+4.3}_{-2.6}\times10^{15}$ & 1.62$^{+1.96}_{-1.52}$\\
\hline
\footnotetext[1]{The column density reported here is the beam-averaged column density, assuming an area filling factor of 1. If the true area filling factor is smaller, this value is an upper limit on the actual beam-averaged column density.}
\footnotetext[2]{For the main fits reported here, we fixed the value of $\Delta$v}
\end{tabular}
\label{hyperfine}
\end{table}  

\begin{table}[ht]
\caption{\am\, Rotational Temperatures}
\centering
\begin{tabular}{ccccccc}
\\[0.5ex]
\hline\hline
& & & & & & \\
Source & T$_{\mathrm{Fit-All}}$ & Ortho / Para Ratio & T$_{\mathrm{Ortho}}$ & T$_{\mathrm{Para}}$ & T$_{13,13 - 11,11}$ & T$_{15,15 - 12,12}$  \\ 
              & ( K )  & & (K) & (K)& ( K)  &  ( K ) \\ [0.5ex] & \\
\hline
{M0.25+0.01}  & 330 $\pm$ 7  & 1.1 $\pm$ 0.1 & 350 $\pm$ 16 & 320 $\pm$ 9 & 410 $\pm$ 75 &   \\ 
{M-0.02-0.07} & 410 $\pm$  7 & 1.1 $\pm$ 0.1 & 430 $\pm$ 10 & 380 $\pm$ 10 & 430 $\pm$ 61 & 450 $\pm$  31 \\ 
{M-0.13-0.08 } & 400 $\pm$ 13 & 1.3 $\pm$ 0.1 & 410 $\pm$ 18 & 380 $\pm$ 21 & 360 $\pm$ 47  & 410 $\pm$ 33 \\
\hline
\end{tabular}
\label{temps}
\end{table}
\clearpage
         
\clearpage

\clearpage

\clearpage

\clearpage

\clearpage

\clearpage

\clearpage

\clearpage

\clearpage

\end{document}